\documentclass[12pt]{article}
\usepackage[a4paper,margin=1in]{geometry}
\usepackage[titletoc,title]{appendix}
\usepackage{changepage}
\usepackage[utf8x]{inputenc}
\usepackage{textcomp,marvosym}
\usepackage{fixltx2e}
\usepackage{amsmath,amssymb}
\usepackage{nameref,hyperref}
\usepackage{authblk}
\usepackage{algorithm}
\usepackage{algpseudocode}
\usepackage{dsfont}
\usepackage{xcolor}
\usepackage{enumitem}
\usepackage{caption,setspace}
\usepackage{subcaption}
\usepackage{pdflscape}
\usepackage{graphicx}
\usepackage{amsthm}
\usepackage{amsmath}
\usepackage{placeins}

\usepackage{nameref,hyperref}

\usepackage{listings}

\newcommand{\dat}{\mathcal{D}}
\newcommand{\paramvec}{\boldsymbol{\theta}}
\newcommand{\paramspace}{\boldsymbol{\Theta}}
\newcommand{\hypeparam}{{\boldsymbol{\lambda}}}
\newcommand{\hypeparamspace}{{\boldsymbol{\Lambda}}}
\newcommand{\like}[2]{\mathcal{L}(#1 ; #2)}
\newcommand{\PDF}[1]{\pi(#1)}
\newcommand{\CondPDF}[2]{\pi(#1 \mid #2)}
\newcommand{\CondPDFsub}[3]{\pi_{#3}(#1 \mid #2)}
\newcommand{\CondCDFsub}[3]{F_{#3}(#1 \mid #2)}
\newcommand{\CondCCDFsub}[3]{G_{#3}(#1 \mid #2)}
\newcommand{\Prob}[1]{\mathbb{P}\left(#1\right)}
\newcommand{\CondProb}[2]{\mathbb{P}\left(#1 \, \middle| \, #2\right)}
\newcommand{\Kernel}[2]{q(#1~\mid~#2)}

\newcommand{\bvec}[1]{\mathbf{#1}}
\newcommand{\ind}[2]{\mathds{1}_{#1}\left(#2\right)}

\lstdefinestyle{customr}{
	belowcaptionskip=1\baselineskip,
	breaklines=true,
	frame=L,
	xleftmargin=\parindent,
	language=R,
	showstringspaces=false,
	basicstyle=\footnotesize\ttfamily,
	keywordstyle=\bfseries\color{green!40!black},
	commentstyle=\itshape\color{purple!40!black},
	identifierstyle=\color{blue},
	stringstyle=\color{orange},
	tabsize=2,
	firstnumber=1,
	captionpos=b,
	numbers=left, 
}

\lstdefinestyle{customc}{
	belowcaptionskip=1\baselineskip,
	breaklines=true,
	frame=L,
	xleftmargin=\parindent,
	language=C,
	showstringspaces=false,
	basicstyle=\footnotesize\ttfamily,
	keywordstyle=\bfseries\color{green!40!black},
	commentstyle=\itshape\color{purple!40!black},
	identifierstyle=\color{blue},
	stringstyle=\color{orange},
	tabsize=2,
	firstnumber=1,
	captionpos=b,
	numbers=left, 
}

\begin{document}

\title{Vector operations for accelerating expensive Bayesian computations -- a tutorial guide}

%\runtitle{A Sample Document}
%\thankstext{T1}{Footnote to the title with the ``thankstext'' command.}

\author[1,2]{David~J. Warne\footnote{To whom correspondence should be addressed. E-mail: david.warne@qut.edu.au}}
\author[3]{Scott~A. Sisson}
\author[1,2]{Christopher Drovandi}

\affil[1]{School of Mathematical Sciences, Queensland University of Technology, Brisbane, Queensland 4001, Australia}
\affil[2]{Centre for Data Science, Queensland University of Technology, Brisbane, Queensland 4001, Australia}
\affil[2]{School of Mathematics and Statistics, University of New South Wales, Sydney, Australia}

\maketitle

\begin{abstract}
	Many applications in Bayesian statistics are extremely computationally intensive. However, they are often inherently parallel, making them prime targets for modern massively parallel processors. Multi-core and distributed computing is widely applied in the Bayesian community, however, very little attention has been given to fine-grain parallelisation using single instruction multiple data (SIMD) operations that are available on most modern commodity CPUs and is the basis of GPGPU computing.  
	%Rather, most fine-grain tuning in the literature has centred around general purpose graphics processing units (GPGPUs). Since the effective utilisation of GPGPUs typically requires specialised programming languages, such technologies are not ideal for the wider Bayesian community. 
	In this work, we practically demonstrate, using standard programming libraries, the utility of the SIMD approach for several topical Bayesian applications.
	% In particular, we consider sampling of the prior predictive distribution for approximate Bayesian computation (ABC), 
	%ABC re-calibration, 
	%the computation of Bayesian $p$-values for testing prior weak informativeness, and inference of on a computationally challenging econometrics model. 
	%Through minor code alterations, 
	We show that SIMD can improve the floating point arithmetic performance resulting in up to $6\times$ improvement in serial algorithm performance. Importantly, these improvements are multiplicative to any gains achieved through multi-core processing.
	%Furthermore $4$-way parallel versions using vector operations can lead to almost $19\times$ improvement over a na\"{i}ve serial implementation. 
	We illustrate the potential of SIMD for accelerating Bayesian computations and provide the reader with techniques for exploiting modern massively parallel processing environments using standard tools.
\end{abstract}

\section{Introduction}

Practical applications in Bayesian statistics are computationally challenging since the posterior density is only known up to a normalising constant.
%That is, given data, $\dat_{\text{obs}}$, a theoretical stochastic model with parameter vector, $\paramvec$, from some parameter space, $\paramspace$, a prior PDF, $\PDF{\paramvec}$, and likelihood function, $\like{\paramvec}{\dat_{\text{obs}}}$, the posterior PDF is
%\begin{equation}
%\CondPDF{\paramvec}{\dat_{\text{obs}}} = \frac{\like{\paramvec}{\dat_{\text{obs}}}\PDF{\paramvec}}{\PDF{\dat_{\text{obs}}}},
%\end{equation}
%where the evidence, $\PDF{\dat}$, is typically 
%%computationally 
%analytically
%intractable. 
Therefore, advanced Monte Carlo schemes are often required. Intractable likelihoods further compound these computational burdens~\cite{Sisson2018}. In addition to posterior sampling, there are other computational challenges in Bayesian statistics. For example, it may be necessary to ensure priors are only weakly informative, and selected from a family of priors with a hyperparameter~\cite{evans2011,Gelman2006}. 

The performance of computational methods and computer hardware continues to improve~\cite{Green2015}. Most recent hardware improvements are due to increased parallel processing capacity. Implementation details for Monte Carlo algorithms are essential to fully exploit modern computational resources. Code optimisation for hardware acceleration is standard in the \emph{high performance computing} (HPC) discipline, and Bayesian practitioners can benefit from these techniques~\cite{Gillespie2017}.  

Many Monte Carlo schemes are inherently parallel. For example, likelihood-free methods, such as approximate Bayesian computation (ABC; Sisson et al. 2018) require a large number of independent prior predictive samples that can be executed in parallel. However, more sophisticated samplers, such as Markov chain Monte Carlo (MCMC)~\cite{Green2015,Marjoram2003}, sequential Monte Carlo (SMC)~\cite{DelMoral2006,Sisson2007} and multilevel Monte Carlo (MLMC)~\cite{Jasra2019,Warne2018} require more effort to efficiently parallelise~\cite{Murray2016} due to synchronisation and communication requirements. These overheads can inhibit scalability for multithreading on central processing units (CPUs). 

General purpose graphics processing units (GPGPUs) are highly effective at accelerating advanced Monte Carlo schemes~\cite{Klingbeil2011,Lee2010}. GPGPUs make heavy use of \emph{single instruction multiple data} (SIMD)~\cite{vanderPas2017}, that is, instructions that operate on vectors element-wise. SIMD is also widely available in modern CPUs containing \emph{vector processing units} (VPUs). By exploiting VPUs, CPU implementations can achieve performance boosts comparable with GPGPU implementations~\cite{Lee2010b,Mudalige2016}. Practical understanding of SIMD for modern CPUs is relevant to applied Bayesian analysis. 
We focus on code structures and algorithmic techniques for practitioners to harness VPUs available in most commodity CPUs, and provide example codes using both R and C\footnote{Example code is available from GitHub: \href{https://github.com/davidwarne/Bayesian_SIMD_examples}{https://github.com/davidwarne/Bayesian\_SIMD\_examples}}.

%\subsection{Parallel computing paradigms}

%For more than a decade, the fundamental speed of CPUs, in terms of clock frequency, has not changed significantly~\cite{Ross2008}. Modern CPUs are typically clocked between $2$ GHz and $4$ GHz. Therefore, to increase the computational throughput, CPU architectures are now designed to perform multiple tasks in parallel. For computationally intensive tasks, such as those frequently encountered in Bayesian statistics, understanding how to exploit parallel computing architectures is essential. 
The paradigms of parallelism are \textit{distributed computing}, \textit{multithreading}, \textit{vectorisation}, and \textit{pipelining}~\cite{Trobec2018,vanderPas2017}. Each paradigm has a granularity that refers to the ratio of communication to computation for a parallel workload. We refer to distributed computing and multithreading as coarse-grained, whereas vectorisation and pipelining are fine-grained. 
Most exemplars of fine-grain parallelism in statistics are based on accelerators, such as,  GPGPUs~\cite{Lee2010,Terenin2018}, Intel Xeon Phis~\cite{Hurn2016,Mahani2015}, and custom co-processors~\cite{Zierke2010}. 
However, accelerators are not practical for many practitioners, so we do not discuss them here. Instead we highlight parallelisation strategies for optimal utilisation of CPUs. In particular, we focus on the effective utilisation of vectorisation using SIMD. %can boost the performance of CPU-based software significantly.  

In this paper, we demonstrate the utility of CPU-based SIMD for accelerating Bayesian inference.  
In Section~\ref{sec:guideVec} we introduce the principles of code optimisation and hardware acceleration using SIMD, then practically demonstrate these principles, in Section~\ref{sec:ABC}, through a tutorial that compares optimised R and C implementations  of prior predictive sampling for ABC~\cite{Sisson2018}. Finally, several topical case studies are presented in Section~\ref{sec:case}: the computation of Bayesian $p$-values for prior weak informativity tests~\cite{evans2011}, and parameter inference in econometrics~\cite{Bekaert2015}.
%We investigate: (i) prior predictive sampling for ABC~\cite{Sisson2018,Sunnaker2013}; 
%(ii) Recalibration post-processing of ABC posterior samples~\cite{Rodrigues2018}; 
%(ii) computing Bayesian $p$-values for prior weak informativity tests~\cite{evans2011,Gelman2006,Nott2018}; (iii) and parameter inference in econometrics~\cite{Bekaert2015,South2019,Li2019}. 
%Several topical, compute intensive, case studies are provided.
We highlight algorithmic features that are suited for parallelism and demonstrate modifications for practical guidance. The clear demonstration of performance benefits using SIMD for CPUs is our main contribution. Optimised implementations are provided as supplementary material using the C language, OpenMP standard (version $\geq 4.5$), the Intel Math Kernel Library\footnote{\href{https://software.intel.com/content/www/us/en/develop/tools/performance-libraries.html}{https://software.intel.com/content/www/us/en/develop/tools/performance-libraries.html}} (MKL) (version $\geq 2018$), and the Intel C compiler~\footnote{\href{https://software.intel.com/content/www/us/en/develop/tools/compilers/c-compilers.html}{https://software.intel.com/content/www/us/en/develop/tools/compilers/c-compilers.html}} \texttt{icc} (version $\geq 17.0.1$). Our guidelines are also directly applicable to the Julia language~\cite{Bezanson2017} and to external interfaces to pre-compiled C code, such as Matlab C-MEX and Rcpp combined with RcppXsimd (see Section~\ref{sec:dis}).

%The remainder of this article is organised as follows: In Section~\ref{sec:guideVec}, we introduce the fundamental programming constructs for exploiting SIMD operations. 
%%Where appropriate, example code is provided. 
%In sections~\ref{sec:ABC},
%%\ref{sec:recab}, 
%\ref{sec:wip}, and \ref{sec:BEGEinf}, we present our case studies on prior predictive sampling for ABC, 
%%ABC recalibration, 
%weak informativity tests, and econometrics parameter inference respectively. 
%%In each case study, we present briefly any essential background theory, highlight the opportunities for parallelisation and vectorisation, demonstrate the steps taken to implement and optimise the computational problem. Performance improvement results are presented for two families of CPU architecture, namely, Haswell and Skylake. 
%Section~\ref{sec:dis} provides a discussion of the results and the strengths and weaknesses of SIMD operations from the perspective of applied Bayesian statisticians.

\section{An introduction to code optimisation}
\label{sec:guideVec}

Here, we introduce CPU computing concepts essential for optimisation. We avoid technological details in favour of simple tools and rules-of-thumb for practitioners.

\subsection{Vectorisation with SIMD}

\begin{figure}[h]
	\centering
	\includegraphics[width=0.6\linewidth]{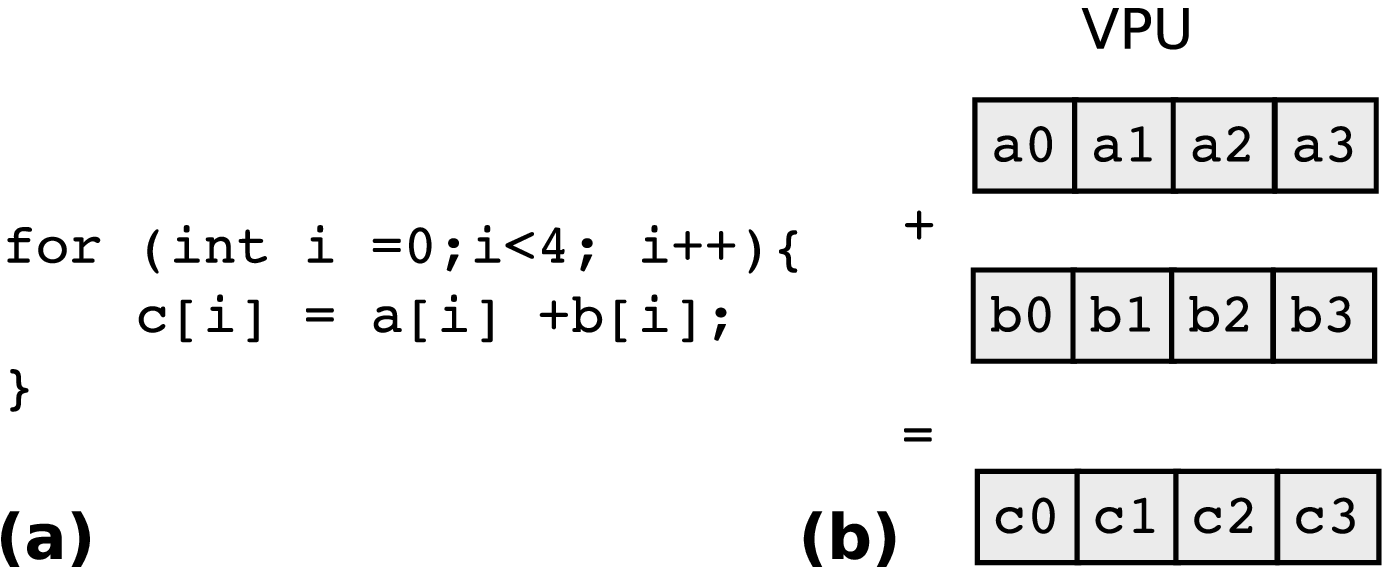}
	\caption{\small VPUs execute in SIMD. The body of the \texttt{for} loop in (a) can be implemented with four scalar additions, or alternatively as one vector addition as shown in (b). The VPUs in this example would need to be 128 bit for single precision floating point inputs and $256$ bit for double precision inputs.}
	\label{fig:vpu}
\end{figure}

CPU cores are sequential\footnote{Although, they do not necessarily execute operations in the order defined by the users code. CPUs re-order operations as needed to optimise pipelining. This is called out-of-order execution and is a feature that is not readily available on accelerator architectures.}, performing a single operation, such as arithmetic or reading/writing data, per clock cycle. Without optimisations, the loop in Figure~\ref{fig:vpu}(a) requires four floating point addition operations.  
Modern CPU cores can perform scalar or vector arithmetic, thus Figure~\ref{fig:vpu}(a) takes only a single \emph{vector} addition ($4 \times$ speedup).

 Vector arithmetic is implemented using VPUs that execute in SIMD. The VPU inputs are wider than the scalar arithmetic units. A VPU that accepts $256$ bit inputs can store eight single precision ($32$ bit) or four double precision ($64$ bit) floating point numbers. VPUs perform arithmetic element-wise on input vectors as shown in Figure~\ref{fig:vpu}(b). 
Acceleration through SIMD is multiplicative with gains achieved through multithreading. Using 18 threads and $512$ bit vectors of an Intel Xeon Gold 6140 CPU~\footnote{\href{https://ark.intel.com/products/120485/Intel-Xeon-Gold-6140-Processor-24-75M-Cache-2-30-GHz-}{https://ark.intel.com/products/120485/Intel-Xeon-Gold-6140-Processor-24-75M-Cache-2-30-GHz-}}, could theoretically perform up to  $18\times (512 / 64) = 144$  times as many double precision operations per second than sequential processing with scalar operations~\cite{Tian2013}.

\subsection{Vectorisation and multithreading with OpenMP}
\begin{figure}[h]
	\centering
	\includegraphics[width=\linewidth]{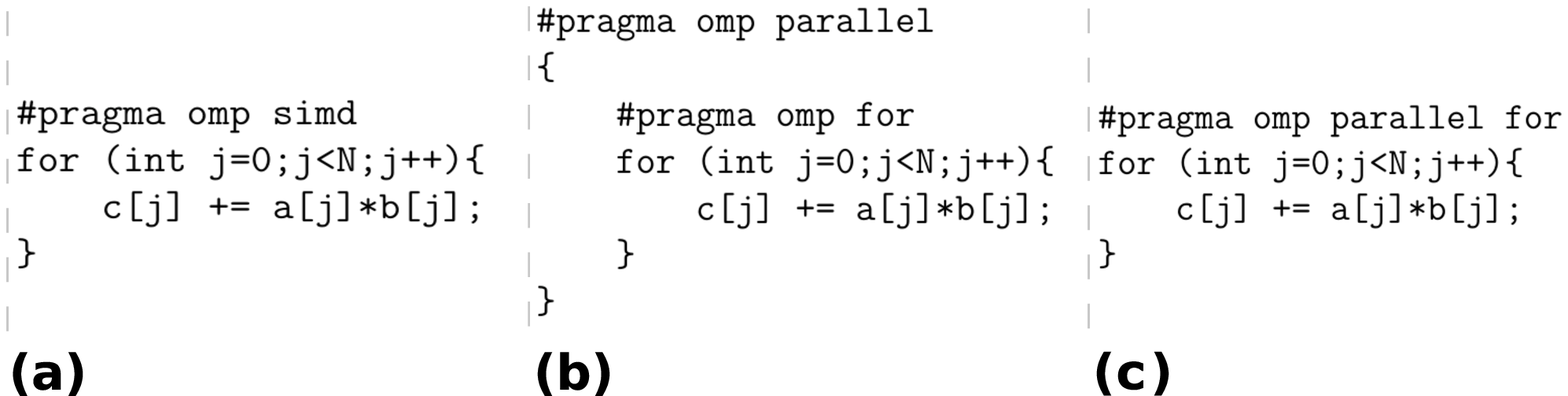}
	\caption{\small Example OpenMP directives:  (a) loop vectorisation, (b) loop parallelisation with multithreading, and (c) shorthand syntax for multithreading.}
	\label{fig:ompsimd}
\end{figure}

 VPUs may be accessed in a variety of ways. A simple approach is OpenMP, an open standard for automated parallel computing. OpenMP supports parallel computing through \emph{directives} within standard C/C++~\cite{vanderPas2017}. The directive \texttt{\#pragma omp simd} (Figure~\ref{fig:ompsimd}(a)) guides the compiler to utilise VPUs. 

 The directive \texttt{\#pragma omp parallel} tells the compiler to create a worker pool of threads within the next code block. Within a parallel block, the directive \texttt{\#pragma omp for} before a loop causes iterations to be distributed across these threads. Figure~\ref{fig:ompsimd}(b) and (c) show the same loop as in Figure~\ref{fig:ompsimd}(a), but parallelised across cores. 
%\begin{figure}[h]
%	\centering
%	\includegraphics[width=0.6\linewidth]{fig3}
%	\caption{\small (a) Loop parallelisation with multithreading. (b) Shorthand syntax.}
%	\label{fig:omppar_for}
%\end{figure} 
 Determining when to use \texttt{simd} (Figure~\ref{fig:ompsimd}(a)),  \texttt{parallel}  and \texttt{for} (Figure~\ref{fig:ompsimd}(b)), and \texttt{parallel for} (Figure~\ref{fig:ompsimd}(c)) is key to performance. The following guidelines are useful:
\begin{enumerate}
	\item Use \texttt{simd} (Figure~\ref{fig:ompsimd}(a)) when loop iterations are independent without conditionals. The loop body should only use arithmetic and standard mathematical functions (e.g.~\texttt{sin}, \texttt{cos}, \texttt{exp} or \texttt{pow}).
	\item  Use \texttt{parallel for} (Figure~\ref{fig:ompsimd}(c)) when loop iterations are independent but conditionals cause iterations to perform different calculations. The loop body can be arbitrarily complex with calls to user defined functions.  
	\item A \texttt{simd} (Figure~\ref{fig:ompsimd}(a)) loop can be included within the body of a \texttt{parallel for} (Figure~\ref{fig:ompsimd}(c)) loop, but the converse is not true. 
	\item The full \texttt{parallel} and \texttt{for} (Figure~\ref{fig:ompsimd}(b)) construct can be useful to more explicitly control the behaviour of individual threads through the use of OpenMP functions.
\end{enumerate}
 The above guidelines are not all strict rules, for example, user defined functions may be called within a \texttt{simd} loop provided the function has been appropriately constructed and the \texttt{declare simd} pre-processor directive is used. Many other features are also available within OpenMP, and we refer the reader to the specifications\footnote{See \url{https://www.openmp.org/specifications/} for full OpenMP specifications.} for details. \cite{Trobec2018} and \cite{vanderPas2017} also provide practical information on parallel computing for HPC. Alternatives to OpenMP are discussed in Section~\ref{sec:dis}.

\subsection{Memory access and alignment}
To fully exploit SIMD, it is crucial to manage memory access. Commodity random access memory (RAM) bandwidth is around $20$ GB/sec, whereas the VPU can process floating point data at a rate of $600$ GB/sec. That means only $3\%$ VPU utilisation is possible for data retrieved from RAM. 
CPUs avoid this bottleneck with a hierarchy of memory caches, typically with three levels: L1, L2, and L3. Lower cache numbers correspond to higher bandwidth, but smaller capacity. L1 cache can keep the VPU close to $100\%$ utilised, but is typically less than $30$ KB. 
When data are requested for computation, the caches are tested in order. If the data are in L1 cache, then there is no memory access penalty. If not in L1 cache, then L2 is checked and so on. RAM is accessed only when the data are not in any cache, which is called a cache miss.
For efficiency, data arrays  should be accessed in patterns that minimise cache misses:
\begin{itemize}
	\item Aim to reuse data soon after it was last accessed. If a partial calculation is pushed out of cache, then it can be faster to re-compute than access the result from RAM.
	\item Avoid random memory access patterns in  favour of regular access \texttt{a[i]}, then \texttt{a[i+n]}, then \texttt{a[i+2*n]}\footnote{Access matrices according to the matrix storage format. For example, the row-major format is used in C/C++ (access row-by-row) and the column-major format is used in R (access column-by-column).}. Ideally, \texttt{n = 1} or is small to avoid cache churn.	
\end{itemize} 

Memory alignment is also important for SIMD. Memory is partitioned into blocks called cache lines. For many recent Intel CPUs, these are $64$ bytes in length\footnote{The CPU-Z utility (\href{https://www.cpuid.com}{https://www.cpuid.com}) is a useful tool to identify cache line sizes.}. When the CPU loads data at an address, the entire cache line is loaded. When doing SIMD, the first byte of a data array should be aligned with the $64$byte cache line boundary, otherwise some vectors will cross cache lines. Performing SIMD on unaligned data incurs a performance penalty, or may fail completely for some older standards. In practice this means:
\begin{itemize}
	\item Use either \texttt{\_mm\_malloc} (Intel) or \texttt{memalign} (GNU), in place of default \texttt{malloc}\footnote{On most systems, \texttt{malloc} defaults is $8$byte or $16$byte.}. 
	\item For static arrays, append \texttt{\_\_declspec(align(n))} to the array declaration\footnote{For example, \texttt{\_\_declspec(align(64)) double A[100]} is a $64$byte aligned array.}. 
\end{itemize}
%For modern processors, SIMD operations can often be applied to unaligned data, but every time a vector crosses a cache line boundary there is a read operation penalty. However, some SIMD operations, particularly from earlier standards, can only operate on aligned data.
Efficient use of cache and memory access patterns is a large topic. See  further discussion in~\cite{Crago2018} and \cite{Geng2018}.

\subsection{Performance analysis}

Code optimisation effort should be informed by performance analysis to ensure reductions in computation time are worth the increase in development time. It is common for this to be overlooked in practice~\cite{Hurn2016,Lee2010b}. Some guidelines to assist are:
\begin{enumerate}
	\item Use profiling and static code analysis software to obtain performance statistics on timing and memory bottlenecks, such as Intel VTune Amplifier\footnote{\href{https://software.intel.com/content/www/us/en/develop/tools/vtune-profiler.html}{https://software.intel.com/content/www/us/en/develop/tools/vtune-profiler.html}} and Intel Advisor\footnote{\href{https://software.intel.com/content/www/us/en/develop/tools/advisor.html}{https://software.intel.com/content/www/us/en/develop/tools/advisor.html}}. These tools are also essential to investigate and control compiler output.
	\item Estimate the theoretical peak performance for your hardware. Compare with an estimate of the  number of floating point operations your algorithm will perform, on average. Use this to estimate the minimum possible runtime~\cite{Hurn2016}. 
	\item Most algorithms require some sequential operations, and not all calculations can be efficiently mapped to SIMD. Amdahl's law~\cite{Amdahl1967} states the speedup factor, $s$, is bounded by
	\begin{equation}
	s \leq \frac{C_S+C_P}{C_S + C_P/P},
	\label{eq:Adl}
	\end{equation}
	where $C_P$ is the sequential runtime of code that can be parallelised, $C_S$ is the runtime of code that must remain sequential, and $P$ is the core count or vector length. Thus, $s = (C_S + C_P)/C_S$ is the maximum speedup as $P \to \infty$.  
\end{enumerate} 

\subsection{A note on random number generation}
When combining statistics with parallel computing it is important to consider initialisation and usage of random number generators (RNGs). Dealing with these aspects poorly can lead to invalid results. In particular, the use of different seed values (one per thread) for the same sequential RNG should be approached with caution. For example, sequences of Linear Congruential Generators can become correlated if a linear sequence of seeds is used~\cite{Davies2007}.

Broadly there are three approaches to deal with this: (i) generate all required randomness serially and distribute among parallel processes; (ii) use a random number generator that can be split into independent sub-streams, for example a generator that supports \emph{Skip Ahead} or \emph{Leapfrog} methods; or (iii) use a sequence of random number generators that are statistically independent for the same seed value. We utilise the last option through the \texttt{VSL\_BRNG\_MT2203} generator family (available within Intel MKL) that provides statistically independent Mersenne Twister generators. 
See~\cite{Bradley2011} and \cite{Lee2010} for details.

\subsection{Summary}

This introduction to vectorisation, multithreading, memory usage, code analysis, and RNGs is not specific to Bayesian statistics. However, the ideas are important to efficiently implement steps within the increasingly computationally expensive algorithms that form a core part of modern practical Bayesian techniques. In Sections~\ref{sec:ABC} and \ref{sec:case} we demonstrate the application of these guidelines through a detailed tutorial and number of case studies in settings of direct interest to Bayesian practitioners.

%\newpage
\section{A practical tutorial demonstration for R users}
\label{sec:ABC}
In this section, we provide a practical demonstration of the computational benefits of directly accessing CPU SIMD operations for ABC-based inference. We begin with R implementations, step through relevant optimisation within R, then demonstrate the computational advantages of direct SIMD access using C and OpenMP. 

\subsection{Prior predictive sampling for approximate Bayesian computation}
ABC techniques are powerful for inference  with intractable likelihood functions~\cite{Sisson2018} and are routinely used to study complex stochastic models~\cite{Ross2017,Tanaka2006}. The most accessible algorithm is \emph{ABC rejection} sampling. Given observed data, $\dat_{\text{obs}}$, the parameter prior distribution, $\pi(\paramvec)$, a discrepancy metric, $\rho$, and  a vector of sufficient (or informative) summary statistics, $S(\dat_{\text{obs}})$, then ABC rejection sampling generates approximate posterior samples by first generating artificial data from the prior predictive distribution,
\begin{equation}
\PDF{\dat} = \int_{\paramspace}s(\dat; \paramvec)\PDF{\paramvec} \, \text{d}\paramvec, \label{eq:ppd}
\end{equation}
where $s(\dat;\paramvec)$ is the probability density of the data generation process for a fixed parameter vector $\paramvec$ in parameter space $\paramspace$. Then a small proportion of samples are accepted to form an approximation to the posterior,
\begin{equation*}
\pi(\paramvec\mid{\mathcal D}_{\text{obs}}) \approx\CondPDF{\paramvec}{\rho(S(\dat_{\text{obs}}),S(\mathcal{D})) \leq \epsilon} = \frac{\CondProb{\rho(S(\dat_{\text{obs}}),S(\mathcal{D})) \leq \epsilon}{\paramvec}\PDF{\paramvec}}{\PDF{\dat_{\text{obs}}}}
\end{equation*} 
 with a sufficently small acceptance threshold $\epsilon$. This procedure is presented in algorithmic form in Algorithm~\ref{alg:ABCrej}.
 
\begin{algorithm}
	\caption{ABC rejection sampling}
	\begin{algorithmic}[1]
		\Repeat
		\State{Generate candidate sample $\paramvec^* \sim \pi(\paramvec)$;}
		\State{Use stochastic simulation to generate prior predictive data $\mathcal{D}^*\sim s(\dat; \paramvec^*)$;}
		\Until{$\rho(S(\dat_{\text{obs}}),S(\mathcal{D}^*)) \leq \epsilon$;}
		\State{Accept $\paramvec^*$ as an approximate posterior sample;}
	\end{algorithmic}
	\label{alg:ABCrej}
\end{algorithm}

In practice, ABC rejection sampling is rarely implemented in this direct manner (Algorithm \ref{alg:ABCrej} is serial, and produces a random number of candidate samples). Rather, it is common to generate a fixed number, $N$, of prior predictive joint samples, $(\paramvec^*,{\mathcal D}^*)$. The acceptance threshold, $\epsilon$, is then selected \emph{a posteriori} based on the empirical distribution of the discrepancy metric, $\rho$. See e.g.~\cite{fan+s18} and \cite{Warne2019} for detailed reviews of Monte Carlo algorithms for ABC.
  ABC-based Monte Carlo estimators  converge slowly in \emph{mean-square} ~\cite{Barber2015}, and therefore require substantial numbers of stochastic model simulations for reliable inference. 

\subsection{Example model: a genetic toggle switch}

%Two example models are selected to demonstrate the advantages and limitations of SIMD parallelism compared with thread parallelism for ABC rejection sampling. Both models are Markov processes: one discrete time; and another continuous time. We find that time continuity, for generating large numbers of prior predictive samples, is the main factor affecting the efficacy of SIMD operations.

%\subsubsection{Genetic toggle switch}
We consider a genetic toggle switch model~\cite{Bonassi2011}. Let $u_i(t) \geq 1$ and $v_i(t) \geq 1$ represent the expression levels of genes $u$ and $v$ at time $t$ for cells $i =1,2 \ldots, C$. Gene expression evolves according to two coupled \emph{stochastic differential equations} (SDEs) 
\begin{align}
\begin{split}
\text{d}u_i(t) &=  \left(\frac{\alpha_u}{1 + v_i(t)^{\beta_u}} - (1 + 0.03 u_i(t))\right)\text{d}t + \frac{1}{2}\text{d}W_{i,u}(t), \\
\text{d}v_i(t) &=  \left(\frac{\alpha_v}{1 + u_i(t)^{\beta_v}} - (1 + 0.03 v_i(t))\right)\text{d}t + \frac{1}{2}\text{d}W_{i,v}(t),
\end{split} \label{eq:gene_uv_SDE}
\end{align}
where  parameters $\alpha_u$, $\beta_u$, $\alpha_v$ and $\beta_v$ define gene inhibition and $W_{i,u}(t)$ and $W_{i,v}(t)$ are independent Wiener processes. The observation process is
\begin{align}
y_i = u_i(T) + \mu + \mu\sigma \frac{\eta_i}{u_{i}(T)^\gamma}, \quad i = 1,2, \ldots, C, \label{eq:gene_u_obs}
\end{align}  
where $T$ is the observation time, $\mu$, $\sigma$ and $\gamma$ control the error rate and $\eta_i$ are standard normal random variables. The data are the measurements for all cells, $\left\{y_i\right\}_{i=1}^C$. Sampling the prior predictive distribution is needed to infer the parameters $\paramvec=(\mu,\sigma,\gamma,\alpha_u, \alpha_v, \beta_u,\beta_v)'$ using ABC. We adopt independent uniform priors as chosen by Bonassi et al. \cite{Bonassi2011}, $\mu \sim \mathcal{U}(250,400)$, $\sigma \sim \mathcal{U}(0.05,0.5)$, $\gamma \sim \mathcal{U}(0.05,0.35)$, $\alpha_u,\alpha_v \sim \mathcal{U}(0,50)$, and $\beta_u,\beta_v \sim \mathcal{U}(0,7)$. We follow Vo et al. \cite{Vo2018} in adopting 19 equally spaced quantiles of the empirical distribution of $\left\{y_i\right\}_{i=1}^C$ as the vector of summary statistics.

\subsection{Implementation and optimisation using R}
To simulate the toggle switch system (Equation~\eqref{eq:gene_uv_SDE}) and observation process (Equation~\eqref{eq:gene_u_obs}) we can use the \emph{Euler-Maruyama} scheme~\cite{Maruyama1955}, 
\begin{align}
\begin{split}
u_i(t+h) &= u_i(t) + \Delta t \frac{\alpha_u}{1 + v_i(t)^{\beta_u}} - \Delta t (1 + 0.03 u_i(t)) + 0.5\sqrt{\Delta t} \xi_{i,u}(t), \\
v_i(t+h) &= v_i(t) + \Delta t \frac{\alpha_v}{1 + u_i(t)^{\beta_v}} - \Delta t (1 + 0.03 v_i(t)) + 0.5\sqrt{\Delta t} \xi_{i,v}(t), 
\end{split} \label{eq:gene_uv_SDE_EM}
\end{align}
where $\Delta t$ is the discretisation step and $\xi_{i,u}(t)$ and $\xi_{i,v}(t)$ are independent normal random variables. For simplicity, we will assume $\Delta t = 1$. 

This can be implemented directly using R as shown in Listing~\ref{lst:naiveR}. Here, the pair of SDEs for $u_i(t)$ and $v_i(t)$ are evolved one cell at a time, for $i = 1,2,\ldots, C$. Such a na\"{i}ve implementation is typical for initial prototypes, but it is profoundly inefficient with a single simulation with $T = 600$ and $C = 8000$ taking $35$ seconds. This is not practical for ABC sampling, even with individual simulations distributed across a HPC cluster.
%\newpage
%\lstset{escapechar=@,style=customr}
\begin{lstlisting}[language=R,style=customr,caption={Na\"{i}ve R implementation of the toggle switch model.},label=lst:naiveR]
simulate.tsw.SDE <- function(theta, T, C) {
	y <- numeric(C); mu <- theta[1]; sigma <- theta[2]; gam <- theta[3]
	alpha.u <- theta[4]; alpha.v <- theta[5]; 
	bet.u <- theta[6]; bet.v <- theta[7]
	for (i in 1:C) { # loop over cells
		ut <- 10; vt <- 10
		for (j in 2:T){ # evolve gene expression dynamics
			p.u <- v.t^bet.u; p.v <- u.t^bet.v
			ut <- 0.97*ut + alpha.u/(1+p.u) - 1.0 + 0.5*rnorm(1,0,1)
			vt <- 0.97*vt + alpha.v/(1+p.v) - 1.0 + 0.5*rnorm(1,0,1)
			ut <- ifelse(ut < 1.0,1.0,u.t)
			vt <- ifelse(vt < 1.0,1.0,v.t)
		}
		# make noise observation
		y[i] <- ut + mu + sigma*mu*rnorm(1,0,1)/(ut^gam)
	}
	y <- ifelse(y < 1.0, 1.0, y)
	return(y)
}
\end{lstlisting}

One can achieve substantial computational improvements by re-writing the R code in terms of vector and matrix mathematics. This exploits optimised linear algebra libraries such as BLAS and LAPACK that often use SIMD. 
Listing~\ref{lst:optR} is an example of vectorised R code for the toggle switch model. 

 The key changes are: 1) genes of all cells are stored in vectors of length $C$; 2) all Gaussian random variates required for the entire simulation are generated in a single call of \texttt{rnorm()}; 3) all genes are simulated together using R's vector maths functions; and 4) logical indexing is used instead of \texttt{ifelse()} to implement the boundary condition. The improvements are substantial, taking $1.3$ seconds to perform a single simulation with $T = 600$ and $C = 8000$. This is more than $25\times$ improvement, without any multithreading, explicit SIMD operations, or Rcpp code.
See the excellent text by Gillespie and Lovelace~\cite{Gillespie2017} for details.% on why these changes have such a large effect.
\newpage
\begin{lstlisting}[language=R,style=customr,caption={Optimised R implementation of the toggle switch model.},label=lst:optR]]
simulate.tsw.SDE <- function(theta, T, C) {
ut <- numeric(C); vt <- numeric(C)
mu <- theta[1]; sigma <- theta[2]; gam <- theta[3]
alpha.u <- theta[4]; alpha.v <- theta[5]
bet.u <- theta[6]; bet.v <- theta[7]
ut[1:C] <- 10; vt[1:C] <- 10
zeta <- matrix(nrow=C,ncol=2*(T-1)+1) 
zeta[,] <- rnorm(C*(2*(T-1)+1),0,1) #generate all random variates
for (j in 2:T) { # evolve all cells together with vectors
p.u <- vt^bet.u; p.v <- ut^bet.v
ut <- 0.97*ut + alpha.u/(1+p.u) - 1.0 + 0.5*zeta[1:C,2*(j-1)]
vt <- 0.97*vt + alpha.v/(1+p.v) - 1.0 + 0.5*zeta[1:C,2*(j-1) + 1]
ut[ut < 1.0] <- 1.0; vt[vt < 1.0] <- 1.0;
}
# make noise observation
y <- ut + mu+  sigma*mu*zeta[1:C,1]/(ut^gam)
y[y < 1.0] <- 1.0
return(y)
}
\end{lstlisting}

The purpose of comparing Listings~\ref{lst:naiveR} and \ref{lst:optR} is to highlight the importance of optimising serial code. 
 Now that the R implementation is efficient, we can distribute the process across multiple cores using the \texttt{doParallel} package as shown in Listing~\ref{lst:doParallel}. Distributing $N = 8064$ draws from the prior predictive across $16$ cores (Intel Xeon E5-2680v3 processor\footnote{\href{https://ark.intel.com/products/81908/Intel-Xeon-Processor-E5-2680-v3-30M-Cache-2-50-GHz-}{https://ark.intel.com/products/81908/Intel-Xeon-Processor-E5-2680-v3-30M-Cache-2-50-GHz-}}) takes just over $11$ minutes!
%\newpage
\begin{lstlisting}[language=R,style=customr,caption={Multithreaded prior predictive sampling using the R package \texttt{doParallel}.},label=lst:doParallel]]
library(tictoc)
library(doParallel)
source("simulateToggleSwitchSDEopt.R")
# set up problem size
T <- 600; C <- 8000; N <- 8064
# set up cluster
cl <- makeCluster(24)
registerDoParallel(cl)
tic() # Generate prior predictive samples
obs_vals <- foreach(k = 1:N) %dopar%  {
	# sample the prior 
	theta <- runif(7,c(250.0,0.05,0.05,0.0,0.0,0.0,0.0),
					 c(400.0,0.5,0.35,50.0,50.0,7.0,7.0))
	# run model simulation 
	c(theta,simulate.tsw.SDE(theta,T,C))
}
toc() # report timing
stopCluster(cl) # clean up
\end{lstlisting}

 For some applications, optimised R can be fast enough and we do not advocate the use of further optimisation for all Bayesian applications. However, the trade-off between development time and runtime is very application specific. For this example, even using 16 cores, the optimised R code will take around 24 hours to draw $N = 1,000,000$ prior predictive samples, which may be insufficient for ABC inference with small $\epsilon$. To motivate progressing beyond R, note that the indirect access of SIMD through optimised BLAS and LAPACK libraries has limitations in terms of cache utilisation for this application. This is highlighted in the next section. 
%\newpage
\subsection{Optimisation using C and SIMD operations}

Direct access to SIMD through C with OpenMP enables superior memory access patterns that ensure substantially higher utilisation of the VPUs (see Section~\ref{sec:dis} for alternatives). While the optimised C implementations presented here are more complex than the optimised R implementation (Listing~\ref{lst:optR}), they provide a practical example that will help reduce the learning curve for practitioners.

Listing~\ref{lst:cFile} provides the main structure of the C program. Note the inclusion of the header files for the MKL libraries and the OpenMP library; we will point out calls to these libraries as necessary. Definitions of the constants \texttt{VECL} and \texttt{ALIGN} refer to the width of the SIMD vectors and the memory alignment to ensure vectors never cross cache lines (see Section~\ref{sec:guideVec}). The functions \texttt{simulate\_tsw\_SDE} and \texttt{main} represent the C equivalents of the R code Listings~\ref{lst:optR} and \ref{lst:doParallel}. C implementations are given in Listings~\ref{lst:optC} and \ref{lst:ppsC}, the details of which are discussed below. 
%\newpage
%\lstset{escapechar=@,style=customc}
\begin{lstlisting}[language=C,style=customc,caption={Overall structure of the optimised C implemenation.},label=lst:cFile]
/* standard C headers*/
#include <stdio.h> 
#include <math.h>
#include <string.h>
/* Intel headers */
#include <mkl.h> 
#include <mkl_vsl.h>
/* OpenMP header */
#include <omp.h> 

#define VECL 4   /* for AVX2 (256 bit = 4 doubles) */
#define ALIGN 64 /* 64 byte cache lines */

void 
simulate_tsw_SDE(VSLStreamStatePtr stream, 
                  double * restrict theta, int T, int C, 
                  double * restrict zeta, double * restrict y) 
	/*... code for model simulation ... */
}

int 
main(int argc,char **argv) { 
	/*... code for parallel prior predictive sampling ... */
}
\end{lstlisting}

The optimised C implementation of the toggle switch model is given in Listing~\ref{lst:optC}. From an algorithmic perspective this code can be considered a hybrid between the two R implementations (Listings~\ref{lst:naiveR} and \ref{lst:optR}). That is, cells are evolved in small blocks of length \texttt{VECL}, as opposed to individually (Listing~\ref{lst:naiveR}) or all together (Listing~\ref{lst:optR}).
%
%\newpage
%\lstset{escapechar=@,style=customc}
The outer loop iterates over the cell index \texttt{c} in strides of \texttt{VECL} (the E5-2680v3 CPU supports 265 bit vectors, giving \texttt{VECL} = 4 doubles). As in the optimised R code (Listing~\ref{lst:optR}), all the random variates required within each block are pre-computed using the \texttt{vdRngGaussian} function from MKL. In each block there are \texttt{VECL} noisy observations, \texttt{y[c]},\texttt{y[c+1]},$\ldots$,\texttt{y[c+VECL-1]}, to be computed (one per cell). Using SIMD we can evolve the SDE pair associated with the \texttt{y[c]} observation and obtain the others in the same block at the same time. This is done with a second loop over the index \texttt{c2} that represents the position in the SIMD vector.
\begin{lstlisting}[language=C,style=customc,caption={Optimised C implementation of the toggle switch model. OpenMP is used to execute the innermost loop using SIMD. MKL is used for RNGs.},label=lst:optC]
void 
simulate_tsw_SDE(VSLStreamStatePtr s, 
                 double * restrict theta, int T, int C, 
                 double * restrict zeta, double * restrict y) {
	double mu = theta[0], sigma = theta[1], gamma = theta[2];
	double *alpha = theta+3, *beta = theta+5;
	/* process cells in blocks of VECL*/
	for (int c=0;c<C;c+=VECL){
		/* Generate all the random variates for these realisations*/
		vdRngGaussian(VSL_RNG_METHOD_GAUSSIAN_BOXMULLER2,s,2*VECL*T,
						zeta,0.0,1.0);
		/* simulate all trajectories for this block in SIMD*/
		#pragma omp simd aligned(zeta:ALIGN, y:ALIGN) 
		for (int c2=0;c2<VECL;c2++) {
			/* copy parameters to ensure in cache/registers*/
			double _gamma = gamma, _sigma = sigma, _mu = mu; 
			double alpha_u = alpha[0], alpha_v = alpha[1];
			double beta_u = beta[0],  beta_v = beta[1];
			double u_t = 10, v_t = 10;
			/* evolve u/v pairs for this cell  */
			for (int j=1;j<T;j++) {
				double p_u = pow(v_t,beta_u);
				double p_v = pow(u_t,beta_v);
				u_t *= 0.97; u_t += alpha_u/(1.0 + p_u) - 1.0;
				v_t *= 0.97; v_t += alpha_v/(1.0 + p_v) - 1.0;
				double zeta_u = zeta[(j-1)*VECL*2 + c2]; 
				double zeta_v = zeta[(j-1)*VECL*2 + 4 + c2]; 
				u_t += 0.5*zeta_u; u_t  = (u_t >= 1.0) ? u_t : 1.0; 
				v_t += 0.5*zeta_v; v_t  = (v_t >= 1.0) ? v_t : 1.0; 
			}
			/* make noisy observation */ 
			y[c+c2] = u_t + _mu + 
			        _sigma*_mu*zeta[(T-1)*VECL*2+c2]/pow(u_t,_gamma);
			y[c+c2] = (y[c+c2] >= 1.0) ? y[c+c2]: 1.0;
		}
	}
}
\end{lstlisting} 
    Note the use of the \texttt{aligned} statement within the \texttt{simd} directive. This allows the compiler to assume the arrays \texttt{zeta} (random variates) and \texttt{y} (simulated data) are aligned to the cache boundary and enables more efficient machine code to be generated. The SIMD loop is not really iterating over values of \texttt{c2}, but rather synchronously executing each statement within the loop for all values of \texttt{c2} concurrently. This leads to exceptional reuse of L1 cache: since the SDE state variable \texttt{u\_t} and \texttt{v\_t} are updated in-place, all computation is performed using the fast CPU memory. 

It is possible to do even better, since computing all Gaussian random variates for each block represents a trade-off. To understand this, note that advanced interfaces, such as compiler intrinsic functions (see discussion in Section~\ref{sec:dis}), can enable efficient RNG sampling within the rest of the Euler-Maruyama update. However, this cannot be done with OpenMP as the function \texttt{vdRngGaussian} has overheads that dominate for very small sample sizes. Conversely, generating all the Gaussian random variates at the start with a single \texttt{vdRngGaussian} call forces the CPU to access L2 cache after a few blocks have been processed, resulting in loss of performance. This trade-off to memory access presented in Listing~\ref{lst:optC} cannot be efficiently replicated in R. Using BLAS and LAPACK functions for small vectors results in similar performance issues as for the \texttt{vdRngGaussian} function. Therefore, the best that can be done with R is to use long vectors to evolve all cells together.

%\newpage
%\lstset{escapechar=@,style=customc}
\begin{lstlisting}[language=C,style=customc,caption={Multithreaded prior predictive sampling using C and OpenMP threading},label=lst:ppsC]
int 
main(int argc,char **argv) { 
	/* set up problem size*/
	int T = 600, C = 8000, K = 7, N = 8064, seed = 1337; 
	/*allocate aligned memory for generated data */
	double *obs_vals = (double *)_mm_malloc(C*sN*sizeof(double),ALIGN);
	/* allocate aligned memory prior samples */
	double *theta = (double *)_mm_malloc(K*N*sizeof(double),ALIGN); 
	/* compute simulations in parallel*/
	#pragma omp parallel shared(seed,N,C,obs_vals,theta)
	{
		VSLStreamStatePtr s;
		/* get thread information and assign workload */
		int tid = omp_get_thread_num();
		int N_per_thr = N/omp_get_num_threads(); 
		/* initialise RNG stream for this thread */
		vslNewStream(&s,VSL_BRNG_MT2203+tid,seed);
		/*allocate aligned memory for Gaussian random variates */
		double *zeta = (double *)_mm_malloc(2*VECL*T*sizeof(double),ALIGN);
		/* compute simulations in this threads workload */
		for (int k=tid*N_per_thr;k<(tid+1)*N_per_thr;k++) {
			/* pointers to memory for this sample*/
			double *p = theta + k*7, *D = obs_vals + k*C;
			/* sample prior */
			vdRngUniform(VSL_RNG_METHOD_UNIFORM_STD,s,1,p,250.0,400.0);
			vdRngUniform(VSL_RNG_METHOD_UNIFORM_STD,s,1,p+1,0.05,0.5);
			vdRngUniform(VSL_RNG_METHOD_UNIFORM_STD,s,1,p+2,0.05,0.35);
			vdRngUniform(VSL_RNG_METHOD_UNIFORM_STD,s,2,p+3,0.0,50.0);
			vdRngUniform(VSL_RNG_METHOD_UNIFORM_STD,s,2,p+5,0.0,7.0);
			/* run simulation and store observations */
			simulate_tsw_SDE(s,p,T,C,zeta,D);
		}
		/* clean up memory */
		vslDeleteStream(&s);
		_mm_free(zeta);
	}
	/* ... code to output results to file goes here ... */
	exit(0);
}
\end{lstlisting}

Armed with a highly efficient SIMD implementation of the model, we can generate prior predictive samples across multiple CPU cores. This uses the multithreading pre-processor directive, \texttt{\#pragma omp parallel}, as shown in Listing~\ref{lst:ppsC}. Each thread is provided with its own random number stream using the Intel MKL/VSL function \texttt{vslNewStream} with independence guarenteed through the  \texttt{VSL\_BRNG\_MT2203} generator family (see Section~\ref{sec:guideVec}). The memory allocated to store prior samples, \texttt{theta}, and simulation outputs, \texttt{obs\_vals}, are shared across threads, using the \texttt{shared} clause, and thread specific offsets are computed, using the OpenMP functions \texttt{omp\_get\_thread\_num()} and  \texttt{omp\_get\_num\_threads()}. The loop over index \texttt{k} is performing the prior predictive sampling tasks assigned to each thread. Unlike the R implementation (Listing~\ref{lst:doParallel}), we explicitly partition the work amongst the available threads to allow reuse of the memory allocated for the Gaussian random variates, \texttt{zeta}. OpenMP does support various forms of automated scheduling of thread tasks, but for this example explicit partitioning is more straight forward.

\subsection{Benchmarks}
\label{sec:bench}
We compare the R implementations (Listings~\ref{lst:naiveR}, ~\ref{lst:optR} and \ref{lst:doParallel}) with the C implementation (Listings~\ref{lst:cFile}, \ref{lst:optC} and \ref{lst:ppsC}) in terms of the time taken to generate a fixed number, $N$, of prior predictive samples given $P$ cores. The fastest sampler would, in turn, produce a smaller Monte Carlo error for a fixed computational budget. For example, the standard central limit theorem suggests a $4\times$ speedup will result in roughly $1/2$ the Monte Carlo error for the same computational budget. In an ABC setting there is also bias to consider, so reducing the model simulation time is absolutely crucial since the acceptance threshold $\epsilon$ must be sufficiently small.

The codes are benchmarked using two Intel Xeon E5-2680v3 (Haswell) processors or a single Intel Xeon Gold 6140 (Skylake) processor. The E5-2680v3 supports the AVX2 instruction set (256 bit vector) and the 6140 supports the AVX512 instruction set (512 bit vectors). Both processor architectures have similar serial performance with cores clocked at around 2.5 GHz. We benchmark with different numbers of cores, $P = 1, 2,4,8,16$ for each implementation. For each value of $P$ the benchmark is repeated four times, to obtain mean computation times $\hat{C}_T$, with each replicate generating $N = 8064$ prior predictive samples. 
For fairness, we compiled R (version 3.3.1) from source using the Intel compiler suite and linked against MKL for optimised BLAS and LAPACK routines. The Intel C compiler version is 17.0.1 (compatible with the GNU C complier version 6.3.0). To demonstrate the difference between speed-up obtained from the improved memory utilisation and the usage of SIMD, we also compile a version of the C implementation with SIMD disabled using the \texttt{-no-vec} compiler option.  

\begin{figure}[h!]
	\includegraphics[width=0.8\textwidth]{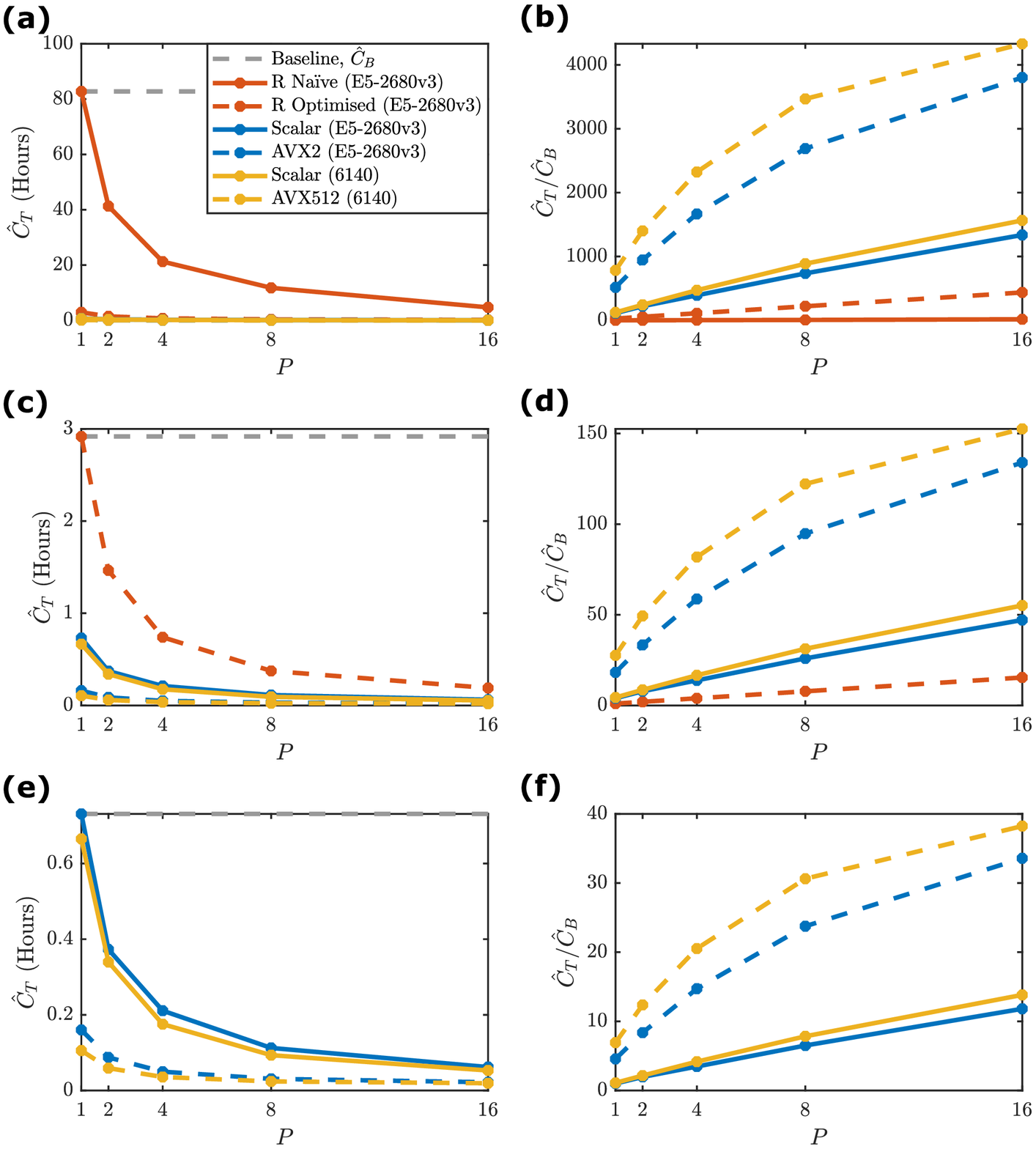}
	\caption{Benchmarking results for the prior predictive sampling application. (a) Mean runtimes, $\hat{C}_T$, using $P$ cores for various implementations: R Na\"{i}ve (Xeon E5-2680v3; solid red), R optimised (Xeon E5-2680v3; dashed red), C with scalar operations only (Xeon E5-2680v3; blue solid), C with AVX2 vectors (Xeon E5-2680v3; blue dashed), C with scalar operations only (Xeon Gold 6140; yellow solid), and C with AVX512 vectors (Xeon Gold 6140; yellow dashed). (b) Computational improvements for each optimisation relative to baseline $\hat{C}_B$ (R Na\"{i}ve with $P=1$). (c) and (d) same as (a) and (b) but using a different baseline  $\hat{C}_B$ (R Optimised with $P = 1$). (e) and (f) same as (a) and (b) but using a different baseline  $\hat{C}_B$ (C with scalar operations only with $P = 1$ on Xeon E5-2680v3).}
	\label{fig:abc_bench}
\end{figure}

Figure~\ref{fig:abc_bench} summarises the results in terms of runtimes $\hat{C}_T$ (Figure~\ref{fig:abc_bench}(a),(c),(e)) and speed-up factor $\hat{C}_T/\hat{C}_B$ for various serial baselines $\hat{C}_B$ (Figure~\ref{fig:abc_bench}(b),(d),(f)). See the supplementary material for precise data tables of runtimes. Using the na\"{i}ve R code as the baseline (Figure~\ref{fig:abc_bench}(a)--(b)), a speedup of $16\times$ is possible with $P = 16$ as one would expect. However, this pales by comparison with the more than $430\times$ speedup using the optimised R code and $P = 16$, demonstrating the value in optimising serial code before implementing parallelism. The improvement is even greater for the C code with speedups of more than $1000\times$ (no vectors, $P = 16$), $3800\times$ (AVX2 vectors, $P = 16$), and $4300\times$ (AVX512, $P = 16$). These impressive numbers are against a na\"{i}ve baseline and more sensible numbers are obtained using the optimised R code as a baseline (Figure~\ref{fig:abc_bench}(c)--(d)). Here optimised C code can achieve $130\times$ speedup (AVX2, $P=16$) and $150\times$ speedup (AVX512, $P=16$) compared with only $16\times$ speedup with optimised R code and $P = 16$.
 Finally, even with the C code as a baseline (Figure~\ref{fig:abc_bench}(e)--(f)), SIMD enables $30\times$ speedup (AVX2, $P = 16$) and almost $40\times$ speedup (AVX512, $P=16$). In this case, the speedup from multithreading alone is only $13\times$ for $P = 16$. There are a few reasons that could cause this, but it is likely that each thread does not perform enough work to mask threading overheads. $P \leq 4$, seems to provide a large enough workload per thread leading to  $15 \times$ speedup (AVX2, $P = 4$) and $20 \times$ speedup (AVX512, $P=4$).

\subsection{Summary}

This tutorial has demonstrated the computational benefits of direct access to SIMD operations in comparison to indirect access through pre-compiled BLAS and LAPACK libraries available in languages such as R and Matlab. Our C implementation (Listings~\ref{lst:optC} and \ref{lst:ppsC}) is more that $20 \times$ faster than our optimised R implementation (Listing~\ref{lst:optR} and \ref{lst:doParallel}) for the same CPU hardware and number of available cores, $P$.  
Restricting comparison to the C implementation, SIMD operations can improve the performance of prior predictive sampling for ABC by a factor of more than $6\times$ compared with standard scalar implementations. Given the enormous number of prior predictive samples required to obtain meaningful parameter estimates with ABC~\cite{Barber2015}, we submit that these techniques are highly relevant for practitioners dealing with such applications. All codes presented here are available as supplementary material.   
\section{Case studies}
\label{sec:case}
In this section, we explore the benefit of SIMD operations in combination with mutlithreading using two case studies relevant to Bayesian practitioners. 
\subsection{Case study 1: weakly informative priors}
\label{sec:wip}
Here, we consider the selection of weakly informative priors~\cite{Gelman2006} within the framework proposed by Evans and Jang~\cite{evans2011} using the implementation by Nott et al. \cite{Nott2018}.
Following Nott et al. \cite{Nott2018}, the prior predictive $p$-value, 
\begin{equation}
p(\dat_{\text{obs}}) = \Prob{\frac{1}{\PDF{\dat}} \geq \frac{1}{\PDF{\dat_{\text{obs}}}}} = \Prob{\PDF{\dat} \leq \PDF{\dat_{\text{obs}}}},
\label{eq:pppv}
\end{equation}
is a measure of Bayesian model criticism,
where $\dat_{\text{obs}}$ are the observational data, and $\PDF{\dat}$ is the prior predictive distribution (Equation~\eqref{eq:ppd}). Equation~\eqref{eq:pppv} provides a $p$-value for prior-data conflict \cite{evans2011} associated with low prior density assigned to parameters with good model fit. If $\gamma$ is a pre-defined small cut-off, then $p(\dat_{\text{obs}}) \leq \gamma$ signifies prior-data conflict. 

Consider a family of priors, $\CondPDF{\paramvec}{\hypeparam}$ with the hyperparameter $\hypeparam \in \hypeparamspace$.
% where $\hypeparamspace$ is the hyperparameter space. 
A base prior, $\CondPDF{\paramvec}{\hypeparam_0}$ for $\hypeparam_0 \in \hypeparamspace$, represents the current best knowledge of the parameter $\paramvec$. Let
\begin{equation}
\CondPDF{\dat}{\hypeparam} = \int_{\paramspace}\like{\paramvec}{\dat}\CondPDF{\paramvec}{\hypeparam}\, \text{d}\paramvec \label{eq:ppdhyp}
\end{equation} 
be the prior predictive distribution for the prior $\CondPDF{\paramvec}{\hypeparam}$ and likelihood $\like{\paramvec}{\dat}$, and let $p_\hypeparam(\dat_{\text{obs}})$ be the prior predictive $p$-value under $\CondPDF{\dat}{\hypeparam}$ (equivalent to Equation~\eqref{eq:pppv}),
\begin{equation}
p_\hypeparam(\dat_{\text{obs}}) = \Prob{\CondPDF{\dat}{\hypeparam} \leq \CondPDF{\dat_{\text{obs}}}{\hypeparam}}.
\label{eq:pppvhype}
\end{equation}

Assume data are generated under the base prior predictive distribution, that is, $\dat_0~\sim~\CondPDF{\dat}{\hypeparam_0}$. The task is to find values of $\hypeparam$ such that,
\begin{equation}
\Prob{p_\hypeparam(\dat_0) \leq \gamma} < \Prob{p_{\hypeparam_0}(\dat_0) \leq \gamma}. \label{eq:weakinfprior}
\end{equation}
Priors satisfying Equation~\eqref{eq:weakinfprior} are weakly informative with respect to the base prior. Weak informativity indicates less prior-data conflict than the base. Equivalently, weak informativity at level $\alpha$ indicates $\gamma_{\hypeparam,\alpha} > \gamma_{\hypeparam_0,\alpha}$ where $\gamma_{\hypeparam,\alpha}$ and $\gamma_{\hypeparam_0,\alpha}$ are the prior-data conflict cut-offs such that  $\Prob{p_\hypeparam(\dat_0) \leq \gamma_{\hypeparam,\alpha}} = \Prob{p_{\hypeparam_0}(\dat_0) \leq \gamma_{\hypeparam_0,\alpha}} = \alpha$.

Given a set of $K$ hyperparameter values, $\hypeparam_1, \ldots, \hypeparam_K \in \hypeparamspace$, with $\hypeparam_0$ corresponding to the base prior, the task is to compute $\gamma_{\hypeparam_k,\alpha}$ such that $\Prob{p_{\hypeparam_k}(\dat_0) \leq \gamma_{\hypeparam_k,\alpha}} = \alpha$ for $k = 0,1,\ldots, K$ and $\dat_0 \sim \CondPDF{\dat}{\hypeparam_0}$. The process proceeds as in Algorithm~\ref{alg:wip}. The output of Algorithm~\ref{alg:wip} is a set of $\alpha$ level cutoffs $\{\gamma_{\hypeparam_0,\alpha},\gamma_{\hypeparam_1,\alpha},\ldots,\gamma_{\hypeparam_K,\alpha}\}$, to derive a set of weakly informative prior distributions with respect to the base prior $\CondPDF{\paramvec}{\hypeparam_0}$, that is, $\{\CondPDF{\paramvec}{\hypeparam_k} :\gamma_{\hypeparam_k,\alpha} > \gamma_{\hypeparam_0,\alpha}, k = 1,2,\ldots,K \}$.

\begin{algorithm}
	\caption{Weak informativity test }
	\begin{algorithmic}[1]
		\State{Initialise hyperparameters $\hypeparam_0, \hypeparam_1, \ldots, \hypeparam_K \in \hypeparamspace$, with base parameter $\hypeparam_0$;}
		     \For{$k \in [0,1, \ldots,K]$}
		          \For{$i \in [1,2,\ldots,N]$}
		               \State{Generate data from base prior predictive $\dat_0^{(i)} \sim \CondPDF{\dat}{\hypeparam_0}$;}
		                \State{Generate data from prior predictive $\dat_k^{(i)} \sim \CondPDF{\dat}{\hypeparam_k}$;}
		                \State{Evaluate $\pi_0^i \leftarrow \CondPDF{\dat_0^{(i)}}{\hypeparam_k}$ and $\pi_k^i \leftarrow \CondPDF{\dat_k^{(i)}}{\hypeparam_k}$;}
		            \EndFor
		            \For{$i \in [1,2,\ldots,N]$}
		                \State{Estimate $p$-value samples $p_k^i \leftarrow p_{\hypeparam_k}(\dat_0^{(i)}) \approx \frac{1}{N}\sum_{j=1}^N \ind{[0,\pi_0^i]}{\pi_k^j}$;}
		             \EndFor
		             \State{Compute $\gamma_{\hypeparam_k,\alpha}$ as the $\alpha$ quantile of $\left\{p_k^1,p_k^2,\ldots, p_k^N\right\}$;} 
		      \EndFor
	\end{algorithmic}
	\label{alg:wip}
\end{algorithm} 

This process is extremely computationally intensive since the evidence term, $\CondPDF{\dat}{\hypeparam}$, must be evaluated for many different $\dat$. Adaptive sequential Monte Carlo (SMC) sampling (Algorithm~\ref{alg:SMC}) using likelihood annealing and Markov chain Monte Carlo (MCMC) proposals~\cite{Beskos2016,Chopin2002} is used to estimate $\CondPDF{\dat}{\hypeparam}$. Algorithm~\ref{alg:wip} requires $2KN$ executions of the adaptive SMC (Algorithm~\ref{alg:SMC}) with $N_p$ particles with tuning parameters $c$, related to the number of MCMC iterations, and $h$, related to the scaling of random walk MCMC proposals. 
\begin{algorithm}[h!]
	\caption{SMC sampler using $N_p$ particles for estimating $\CondPDF{\dat}{\hypeparam}$ 
		% where $\hypeparam$ with user defined values of $c$ and $h$ 
	}
	\begin{algorithmic}[1]
		\State{Initialise $n = 0$, $t_0 = 0$, $\CondPDFsub{\dat}{\hypeparam}{0} = 1$, $\paramvec_0^{(i)} \sim \CondPDF{\paramvec}{\hypeparam}$, and $W_0^{(i)} = 1/N_p$ for $i = 1,2,\ldots, N_p$, and specify user-defined values for tuning parameters $h$ and $c$;}
		\Repeat
		\State{Set $n \leftarrow n + 1$;}
		\State{Find $t$ such that  $\left[\sum_{i=1}^{N_p} \left(\frac{W_{n-1}^{(i)}\like{\paramvec_{n-1}^{(i)}}{\dat}^{t_n - t_{n-1}}}{\sum_{j=1}^{N_p} W_{n-1}^{(j)}\like{\paramvec_{n-1}^{(i)}}{\dat}^{t_n - t_{n-1}}}\right)^2\right]^{-1} = N_p/2$, that is, the effective sample size (ESS) is $N_p/2$; Set $t_n \leftarrow \min (1,t)$;}
		
		\State{Set $W_{n}^{(i)} \leftarrow W_{n-1}^{(i)}\like{\paramvec_{n-1}^{(i)}}{\dat}^{t_n - t_{n-1}}$ for $i = 1,2,\ldots, N_p$;}
		\State{Set $\CondPDFsub{\dat}{\hypeparam}{n} = \CondPDFsub{\dat}{\hypeparam}{n-1}\left[\frac{1}{N_p}\sum_{i=1}^{N_p}\like{\paramvec_{n-1}^{(i)}}{\dat}^{t_n - t_{n-1}}\right]$}
		\State{Resample with replacement the particles $\paramvec_n^{(i)}\sim \{\paramvec_n^{(j)},W_n^{(j)}\}_{j=1}^{N_p}$ for $i=1,2,\ldots,N_p$;}
		\State{Set $W_{n}^{(i)} \leftarrow 1/N_p$ for $i=1,2,\ldots,N_p$;}
		\State{Construct a tuned  proposal kernel $\Kernel{u}{v}~=~\phi(u ; v, h^2\hat{\Sigma}_n)$, where $\hat{\Sigma}_n$ is the sample covariance matrix of $\{\paramvec_n^{(j)}\}_{j=1}^{N_p}$; }
		\State{Set $R_n \leftarrow \lceil \frac{\ln c}{\ln (1-p_{\text{acc}})} \rceil$ where $p_{\text{acc}}$ is the estimated acceptance probability determined from initial trial MCMC iterations and $\lceil\cdot \rceil$ is the ceiling function;}
		\For{$i\in [1,2\ldots,N_p]$} \label{line:MCMCstart}
		\For{$r=1,2,\ldots,R_n$}
		\State{Generate proposal $\paramvec^* \sim \Kernel{\paramvec}{\paramvec_{n}^{(i)}}$}
		\State{Compute acceptance ratio, $\alpha(\paramvec^*,\paramvec_{n}^{(i)}) \leftarrow \frac{\like{\paramvec^*}{\dat}^{t_n}\CondPDF{\paramvec^*}{\hypeparam} \Kernel{\paramvec_{n}^{(i)}}{\paramvec^*}}{\like{\paramvec_n^{(i)}}{\dat}^{t_n}\CondPDF{\paramvec_n^{(i)}}{\hypeparam}\Kernel{\paramvec^*}{\paramvec_n^{(i)}}}$ }
		\State{With probability $\min\left(1,\alpha(\paramvec^*,\paramvec_{n}^{(i)})\right)$, set $\paramvec_n^{(i)} \leftarrow \paramvec^*$;}
		\EndFor
		\EndFor \label{line:MCMCend}
		\Until{$t_n = 1$ }
	\end{algorithmic}
	\label{alg:SMC}
\end{algorithm}

We consider weakly informative priors for the analysis of an acute toxicity test in which $M$ groups of animals are given different dosages of some toxin, and the number of deaths in each group are recorded~\cite{evans2011,Nott2018}. A logistic regression model is applied for the number of deaths $y_i$ in group $i \in [1,2,\ldots M]$,
\begin{equation}
y_i \sim \text{Bin}\left(n_i,\frac{1}{1 + e^{\beta_0 + \beta_1 x_i}}\right), \quad i = 1,2,\ldots, M,
\label{eq:bioassay}
\end{equation}
where $n_i$ and $x_i$ are respectively the number of animals and the toxin dose level in the $i$th group of the experiment, and $\paramvec = (\beta_0,\beta_1)'$ are the regression parameters. Given data $\dat_{\text{obs}} = [y_{\text{obs},1},y_{\text{obs},2},\ldots,y_{\text{obs},M}]$, the likelihood function is 
\begin{equation}
\like{\paramvec}{\dat_{\text{obs}}} = \prod_{i=1}^M \binom{n_i}{y_{\text{obs},i}}\left(\frac{1}{1 + e^{\beta_0 + \beta_1 x_i}}\right)^{y_{\text{obs},i}}\left(1-\frac{1}{1 + e^{\beta_0 + \beta_1 x_i}}\right)^{n_i-y_{\text{obs},i}}.
\end{equation}
From the data in \cite{evans2011} we have $M = 4$, $n_1 = n_2 = n_3 = n_4 = 5$, $x_1 = -0.86$, $x_2 = -0.3$, $x_3 = -0.05$ and $x_4 = -0.75$.
We adopt bivariate Gaussian priors from \cite{Nott2018}, $\paramvec~\sim~\mathcal{N}(\bvec{0},\text{diag}(\hypeparam)^2)$ and $\hypeparam_0 = (10,2.5)'$.

\subsubsection{Parallelisation and SIMD opportunities}

Parallel implementations of SMC samplers using multithreading require thread synchronisation for both the resampling and annealing steps~\cite{Hurn2016,Lee2010,Murray2016}. Thus, it is more beneficial to distribute the $K$ hyperparameter values in Algorithm~\ref{alg:wip} across $P$ cores, with each thread computing the $\alpha$ quantile for $K/P$ hyperparameters, since these may be performed independently. For every hyperparameter, we sequentially process the $N$ $p$-value computations. The SMC sampler (Algorithm~\ref{alg:SMC}) can be accelerated through SIMD.
SMC samplers are well suited to SIMD since sychronisation is maintained automatically. While there are many aspects of Algorithm~\ref{alg:SMC} that use SIMD in the code example, we focus here on SIMD for the MCMC proposal kernel for diversification of particles (Steps~\ref{line:MCMCstart}--\ref{line:MCMCend} of Algorithm~\ref{alg:SMC}) as it has the greatest effect on performance.

The strategy is similar to the C implementation of the toggle switch model (Section~\ref{sec:ABC}). At SMC step $n$, each particle is perturbed via $R_n$ MCMC steps. The same operations are performed at each MCMC iteration, with the exception of a single branch operation that arises from the accept/reject step. We process the particle updates in blocks of length $V$ and evolve the $R_n$ MCMC steps for this block together using SIMD. All of the Gaussian and uniform random variates required for the block are generated together before the block is processed. %We use the same vector notation as in %Section~\ref{sec:ABC}, however, we also define  
The SIMD MCMC proposals are performed as in Algorithm~\ref{alg:MCMCvec}.
\begin{algorithm}
	\caption{SIMD implementation of MCMC proposals for SMC}
	\begin{algorithmic}[1]
		\For{$i = 1,1+V, 1+2V,\ldots,N_p+1-V, N_p$}
		\State{Generate increments $\xi_{1:V}^{1:R_n} \sim \mathcal{N}(\bvec{0},h\hat{\Sigma})$;}  
		\State{Generate uniform variates $u_{1:V}^{1:R_n} \sim \mathcal{U}(0,1)$;} 
		\For{$r=1,2,\ldots,R_n$}
		\State{Generate proposal $\paramvec_{1:V}^* \leftarrow \paramvec_n^{(i:i+V)} + \xi_{1:V}^{r}$;}
		\State{ $\alpha_{1:V} \leftarrow \mathcal{L}^V(\paramvec_{1:V}^*; \dat)^{\circ t_n} \circ \pi^V(\paramvec_{1:V}^*| \hypeparam) \circ \phi^V(\paramvec_n^{(i:i+V)}; \paramvec_{1:V}^*,h\hat{\Sigma})$;}
		\State{$\alpha_{1:V} \leftarrow \alpha_{1:V} \oslash \left[\mathcal{L}^V(\paramvec_n^{(i:i+V)}; \dat)^{\circ t_n} \circ \pi^V(\paramvec_n^{(i:i+V)}| \hypeparam) \circ \phi^V(\paramvec_{(i:i+V)}^*; \paramvec_n^{(i:i+V)},h\hat{\Sigma})\right]$;}
		\For{$j = 1,2,\ldots, V$}
		\If{$u_{j}^r \leq \alpha_j$}
		\State{$\paramvec_n^{(i:i+j-1)} \leftarrow \paramvec_{j}^*$;}
		\EndIf
		\EndFor
		\EndFor
		\EndFor
	\end{algorithmic}
	\label{alg:MCMCvec}
\end{algorithm}
 Hadamard notation indicates element-wise division, $x \oslash y$, multiplication, $x \circ y$ and exponentiation, $x^{\circ a}$, for scalar $a$. Element-wise application of a function, $f$, over length $V$ vectors is denoted by $f^{V}(x_{i:i+V}) = [f(x_i),f(x_{i+1}),\ldots, f(x_{i+V})]$. 
In practice, Algorithm~\ref{alg:MCMCvec} works with the likelihood and prior on the log scale to avoid numerical underflow. For optimal performance, SIMD forms for the likelihood $\mathcal{L}^V$, prior density $\pi^V$ and Gaussian proposal density $\phi^V$ are required. This is available for the logistic regression model, Gaussian priors and proposals used here. 
Other aspects of Algorithm~\ref{alg:SMC} that can utilise SIMD include likelihood evaluations and the computation of ESS and weight updates.

The scalar bottleneck is the multinomial resampling step that utilises the look-up method. Parallel approximations for resampling have been proposed~\cite{Murray2016} and demonstrated to be very effective for large scale SMC samplers. Since we focus here on a SIMD implementation of SMC, we do not implement this, however, extending \cite{Murray2016} to SIMD is an important piece of future work.

%\textcolor{red}{[Comment on the last two paragraphs: I think if you are using this paper for people to learn how to vectorise algorithms, then you might want to consider elaborating on the other vectorisable components a little more, perhaps being a little more explicit. Here you just focus on one part and leave the rest as an exercise for the reader, which is ok if your reader is competent, but if they are just learning then maybe it needs more. Depends on how much you want to spell it out I suppose. Just a thought.]}

\subsubsection{Performance}

We test the performance improvement obtained through vectorisation and multithreading using the Xeon E5-2680v3 and Xeon Gold 6140 processors. The evaluation of the weak informativity test in Algorithm~\ref{alg:wip} is performed for $K = 400$ hyperparameters and $N = 400$ datasets. In all simulations, the $K$ hyperparameter values are generated using a bivariate uniform distribution $\hypeparam_k \sim \mathcal{U}([0.1,10]\times [0.1,20])$ for $k = 1,2,\ldots, K$. The SMC sampling is performed with $N_p = 500$ particles, with the lower particle count enabling computation to remain largely within L1 and L2 cache. Tuning parameter values are specified as $c = 0.01$ and $h = 2.38/\sqrt{2}$. Results are provided in Figure~\ref{fig:wip} (see supplementary material for data tables).
\begin{figure}[h!]
	\includegraphics[width=0.8\textwidth]{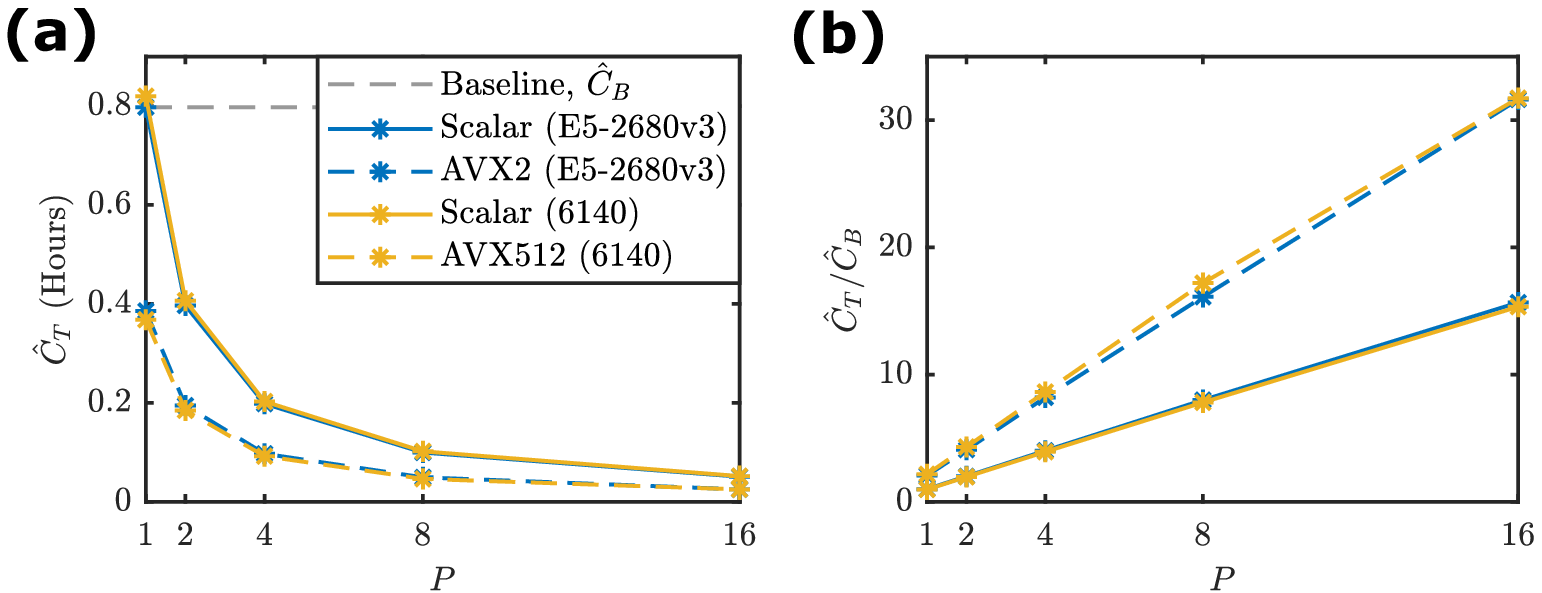}
	\caption{Benchmarking results for weak informativity test. (a) Mean runtimes, $\hat{C}_T$, using $P$ cores for various implementations:  C with scalar operations only (Xeon E5-2680v3; blue solid), C with AVX2 vectors (Xeon E5-2680v3; blue dashed), C with scalar operations only (Xeon Gold 6140; yellow solid), and C with AVX512 vectors (Xeon Gold 6140; yellow dashed). (b) Computational improvements for each optimisation relative to baseline  $\hat{C}_B$ (C with scalar operations only with $P = 1$ on Xeon E5-2680v3).}
	\label{fig:wip}
\end{figure}

%\begin{figure}[h]
%	\includegraphics[width=\textwidth]{./wip_fig}
%	\caption{test}
%	\label{fig:wip}
%\end{figure}

 %Tables~\ref{tab:wip_avx256} and \ref{tab:wip_avx512}.

Figure~\ref{fig:wip}(b) shows almost perfect speedup from multithreading and consistent improvement due to SIMD when comparing scalar and SIMD performance for the same CPU and core count ($2.1 \times$ for AVX2 and $2.3 \times$ for AVX512). Diminishing returns are observed when stepping from $256$ bit vector operations to $512$ bit SIMD.
In our bioassay example, the MCMC proposal kernel performs a small number of steps (usually no more than $R_n = 30$) and the number of particles in the SMC sampler is $N_p = 500$. The performance will improve for SIMD in cases where longer MCMC runs are required, since the MCMC step will dominate SMC iterations and more reuse of L1 cache will occur.

\subsubsection{Summary}

We have presented the more challenging problem of weak informativity tests over a family of priors $\CondPDF{\paramvec}{\hypeparam}$ with respect to a base prior $\CondPDF{\paramvec}{\hypeparam_0}$. This requires a large number of approximations of the posterior normalising constant for different datasets from the prior predictive distribution. The traditional approach of parallelisation of each SMC iteration would reduce the level of parallelism available across hyperparameters. Using the SIMD implementation of SMC  we double our computational performance and can reserve multithreading across hyperparameters. A combination of parallelism and SIMD could be implemented to improve the performance of a single SMC step. However, this offers little benefit here since utilising threads for SMC forces hyperparameters to be processed serially. If HPC resources are available then hyperparameters could be distributed across servers: leaving both threading and SIMD for SMC steps. 

\subsection{Case study 2: Parameter inference for a non-Gaussian asymmetric volatility model}
\label{sec:BEGEinf}
We now consider an adaptive SMC sampler to perform parameter inference in eleven dimensions for the ``bad evironment -- good environment'' (BEGE) model of innovations on stock market returns~\cite{Bekaert2015,South2019}.

\subsubsection{The BEGE model}
The BEGE model \cite{Bekaert2015} is a non-Gaussian generalisation of the Glosten-Jagannathan-Runkle (GJR) asymmetric volatility model~\cite{Glosten1993} and is a generalised autoregressive conditional heteroskedasticity (GARCH) model~\cite{Bollerslev1986}. The BEGE model describes the time-series of stock market returns, $\{r_t\}_{t\geq0}$ using a model on innovation on returns, $\{u_t\}_{t \geq 0}$, that consists of a linear combination of positive ``good environment'' shocks, $\{\omega_{p,t}\}_{t\geq0}$, and negative ``bad environment'' shocks, $\{\omega_{n,t}\}_{t \geq 0}$. The BEGE time-series evolves according to
\begin{align}
\begin{split}
r_{t+1} = u_{t+1} + \mu, &\quad\ u_{t+1} = \sigma_p \omega_{p,t+1} - \sigma_n \omega_{n,t+1},\\
\omega_{p,t+1} \sim \tilde{\Gamma}(p_t,1),&\quad
\omega_{n,t+1} \sim \tilde{\Gamma}(n_t,1),
\end{split}
\label{eq:BEGE}
\end{align}
where $\mu$ is the conditional mean of returns, $\tilde{\Gamma}(k,1)$ is the centered (de-meaned) gamma distribution with shape $k$ and unit scale, $\{p_t\}_{t\geq0}$ and $\{n_t\}_{t\geq0}$ are respectively the shapes of the positive and negative shocks, and $\sigma_p$ and $\sigma_n$ are their constant scales. The shape parameters evolve according to
\begin{align}
\begin{split}
p_t &= p_0 + \rho_p p_{t-1} + \frac{\phi^+_p}{2\sigma_p^2}u_t^2 \ind{[0,\infty)}{u_t} + \frac{\phi^-_p}{2\sigma_p^2}u_t^2\ind{(-\infty,0)}{u_t}, \\
n_t &= n_0 + \rho_n n_{t-1} + \frac{\phi^+_n}{2\sigma_n^2}u_t^2 \ind{[0,\infty)}{u_t} + \frac{\phi^-_n}{2\sigma_n^2}u_t^2\ind{(-\infty,0)}{u_t},
\end{split}\label{eq:shape}
\end{align}
where $p_0$, $n_0$ are initial conditions, and $\rho_p$, $\rho_n$, $\phi^+_p$, $\phi^+_n$, $\phi^-_p$, and $\phi^-_n$ are autoregression parameters.

We use S\&P Composite Index returns over the period July 1926 to January 2018 (obtained from the Center of Research in Security Prices) consisting of $T = 1099$ months of logged monthly divided-adjusted returns, $\dat = R_{\text{obs}, T} = \left\{r_{\text{obs},t}\right\}_{0\leq t \leq T}$. Using these data, Bayesian inference on the unknown parameters $\paramvec = \left(p_0,\sigma_p,\rho_p,\phi_p^+,\phi_p^-,n_0,\sigma_n,\rho_n,\phi_n^+,\phi_n^-,\mu\right)$ is performed. Priors are adopted from~\cite{South2019} and are given by $p_0 \sim \mathcal{U}(10^{-4},0.5)$, $\sigma_p \sim \mathcal{U}(10^{-4},0.3)$, $\rho_p \sim \mathcal{U}(10^{-4},0.99)$, $\phi_p^+ \sim \mathcal{U}(10^{-4},0.5)$, $\phi_p^- \sim \mathcal{U}(10^{-4},0.5)$, $n_0 \sim \mathcal{U}(10^{-4},1)$, $\sigma_n \sim \mathcal{U}(10^{-4},0.3)$, $\rho_n \sim \mathcal{U}(10^{-4},0.99)$, $\phi_n^+ \sim \mathcal{U}(-0.2,0.1)$, $\phi_n^- \sim \mathcal{U}(10^{-4},0.75)$, and $\mu_t \sim \mathcal{U}(-0.9,0.9)$. 

The major challenge in this inference problem is the computational cost associated with the evaluation of the log-likelihood function,
\begin{equation}
\log \like{\paramvec}{\dat} = \log \like{\paramvec}{R_{\text{obs}, T}} =  \sum_{t=1}^T \log \CondPDF{r_{\text{obs},t}}{r_{\text{obs},t-1},\paramvec}. \label{eq:BEGE_LL}
\end{equation}
Bekaert et al. \cite{Bekaert2015} show that evaluation of the transitional densities, $\CondPDF{r_{\text{obs},t}}{r_{\text{obs},t-1},\paramvec}$, requires the so called BEGE density $\CondPDFsub{u_t}{\sigma_p,\sigma_n,p_{t-1},n_{t-1}}{\text{BEGE}}$. This can be approximated by computing the BEGE distribution function and taking finite differences~\cite{Bekaert2015}. The BEGE distribution function is given by
\begin{equation}
\CondCDFsub{u_t}{\sigma_p,\sigma_n,p_{t-1},n_{t-1}}{\text{BEGE}} = \int_{-\infty}^{\infty} \CondCCDFsub{\omega_{p,t} - u_{t}}{n_{t-1},\sigma_n}{\tilde{\Gamma}}\CondPDFsub{\omega_{p,t}}{p_{t-1},\sigma_p}{\tilde{\Gamma}}\, \text{d}\omega_{p,t}, \label{eq:CDF_BEGE}  
\end{equation}
with $\CondCCDFsub{\omega_{p,t} - u_{t}}{n_{t-1},\sigma_n}{\tilde{\Gamma}} = 1 - \CondCDFsub{\omega_{p,t} - u_{t}}{n_{t-1},\sigma_n}{\tilde{\Gamma}}$, 
where $\CondPDFsub{\cdot}{k,s}{\tilde{\Gamma}}$ and $\CondCDFsub{\cdot}{k,s}{\tilde{\Gamma}}$ are, respectively, the probability density and distribution functions of a centered gamma distribution with shape $k$ and scale $s$. The integral in Equation~\eqref{eq:CDF_BEGE} can be approximated numerically using quadrature. For each point in the discretisation of the $\omega_{p,t}$ parameter space, we require two evaluations of the incomplete gamma function,
\begin{equation}
P(a,x) = \frac{1}{\Gamma(a)} \int_{0}^x e^{-t}t^{a-1}\, \text{d}t, \label{eq:incompGamma}
\end{equation}
where $\Gamma(a)$ is the gamma function. Equation \eqref{eq:incompGamma} can be computed using a power series \cite{Press1992}. 

We apply an adaptive SMC sampler to move $N_p$ particles from the prior $\PDF{\paramvec}$ to the posterior $\CondPDF{\paramvec}{R_{\text{obs},T}}$ under the BEGE model. This is a very similar SMC sampler to that applied in Section~\ref{sec:wip} (Algorithm~\ref{alg:SMC}). The main difference is the scaling rule for the proposal kernel within the MCMC step, as the posterior is highly non-Gaussian. We apply the method of ~\cite{Salomone2018} to evaluate a set of MCMC trials each with a random scale factor, $h \in [0.1,0.2, \ldots, 1.0]$, at each SMC iteration. We then choose the scale factor, $h_{\text{opt}}$, which maximises the median expected squared jump distance across all particles. This continues until at least half of the particles have moved further than the median~\cite{Salomone2018}.

\subsubsection{Parallelisation and vectorisation opportunities}

In the BEGE model inference problem, we utilise both SIMD and multithreading to accelerate a single SMC sampler. 
Parallel implementations of SMC samplers and particle filters have been well studied in the literature~\cite{Hurn2016,Lee2010,Murray2016}. Within a single SMC iteration, particles are completely independent of each other. However, as noted in Section~\ref{sec:wip}, automatic synchronisation can occur within the MCMC proposal mechanism. Rather than distribute $N_p$ particles across $P$ cores, we ensure that the distribute occurs in contiguous blocks of length $V$. Each core will process $N_p / (PV)$ blocks of particles. 
All data associated with particles is processed in contiguous blocks of length $V$, allowing each thread to independently exploit SIMD within the MCMC proposal mechanism as per Algorithm~\ref{alg:MCMCvec}. 

Another way to exploit SIMD is in the evaluation of the BEGE log-likelihood. We extend the approximation of the integral in Equation~\eqref{eq:CDF_BEGE}. Consider a discretisation of $\omega_{p,t}$ of $N_\omega+1$ nodes with spacing $\Delta \omega$. Then Equation~\eqref{eq:CDF_BEGE} can be approximated by
\begin{align}
\begin{split}
\CondCDFsub{u_t}{\sigma_p,\sigma_n,p_{t-1},n_{t-1}}{\text{BEGE}} \approx \sum_{j=1}^{N_\omega}\left\{ \left[1 - P\left(n_{t-1},\frac{\omega_{p,t}^{j-1} - u_t + n_{t-1}\sigma_{n}}{\sigma_{n}}\right)\right]\right. \\\times \left. \left[P\left(p_{t-1},\frac{\omega_{p,t}^j + p_{t-1}\sigma_{p}}{\sigma_{p}}\right) -P\left(p_{t-1},\frac{\omega_{p,t}^{j-1} + p_{t-1}\sigma_{p}}{\sigma_{p}}\right)\right]\right\},
\end{split}\label{eq:BEGE_LL_approx}
\end{align}
where $\omega_{p,t}^j = \omega_{p,t}^0 + j\Delta\omega$, for $j = 0,1, \ldots, N_\omega$ and $\omega_{p,t}^0$ is the lower bound of the discretisation. For every point $\{\omega_{p,t}^j\}_{0\leq j\leq N_\omega}$, the incomplete gamma function is evaluated twice; once with $p_{t-1}$ and once with $n_{t-1}$. Therefore, we can consider SIMD for the incomplete gamma function for blocks of $\omega_{p,t}$ points of length $V$, that is, $P^V(a,x_{1:V}) = \left[P(a,x_1),P(a,x_2),\ldots,P(a,x_V)\right]$. We achieve this by extending the method of \cite{Press1992}, resulting in the element-wise vector series expression 
\begin{equation}
\begin{split}
P^V(a,x_{1:V}) =& \left[\left(\exp^V\left(-x_{1:V}\right) \circ (x_{1:V})^{\circ a}\right) \oslash \left(\Gamma(a)\bvec{e}_{1:V}\right)  \right] \\ &\circ \left[\sum_{j=0}^\infty \left((x_{1:V})^{\circ j}\right) \oslash \left((j+1)\bvec{e}_{1:V}\right)\right],
\end{split}\label{eq:IGF_series_vec}
\end{equation}
where $\exp^V\left(-x_{1:V}\right) = \left[e^{-x_1},e^{-x_2},\ldots,e^{-x_V} \right] \in \mathbb{R}^{1\times V}$, and $\bvec{e}_{1:V}~=~[1,1,1,\ldots,1]~\in~\mathbb{R}^{1\times V}$. Equation~\ref{eq:IGF_series_vec} allows efficient iteration that reuses previous steps, thus enabling good use of L1 and L2 cache. The series is truncated once it has converged under the $\infty$-norm. 

We proceed to approximate $\CondCDFsub{u_t}{\sigma_p,\sigma_n,p_{t-1},n_{t-1}}{\text{BEGE}}$ by: 1) applying Equation~\eqref{eq:IGF_series_vec} across the discretisation, $\{\omega_{p,t}^j\}_{0 \leq j \leq N_\omega}$, in blocks of length $V$ with shape $p_{t-1}$; 2) applying Equation~\eqref{eq:IGF_series_vec} across the discretisation, $\{\omega_{p,t}^j\}_{0 \leq j \leq N_\omega}$, in blocks of length $V$ with shape $n_{t-1}$; and 3) accumulating the sum of products in Equation~\ref{eq:BEGE_LL_approx}.

\subsubsection{Performance}

\begin{figure}[h!]
	\includegraphics[width=0.85\textwidth]{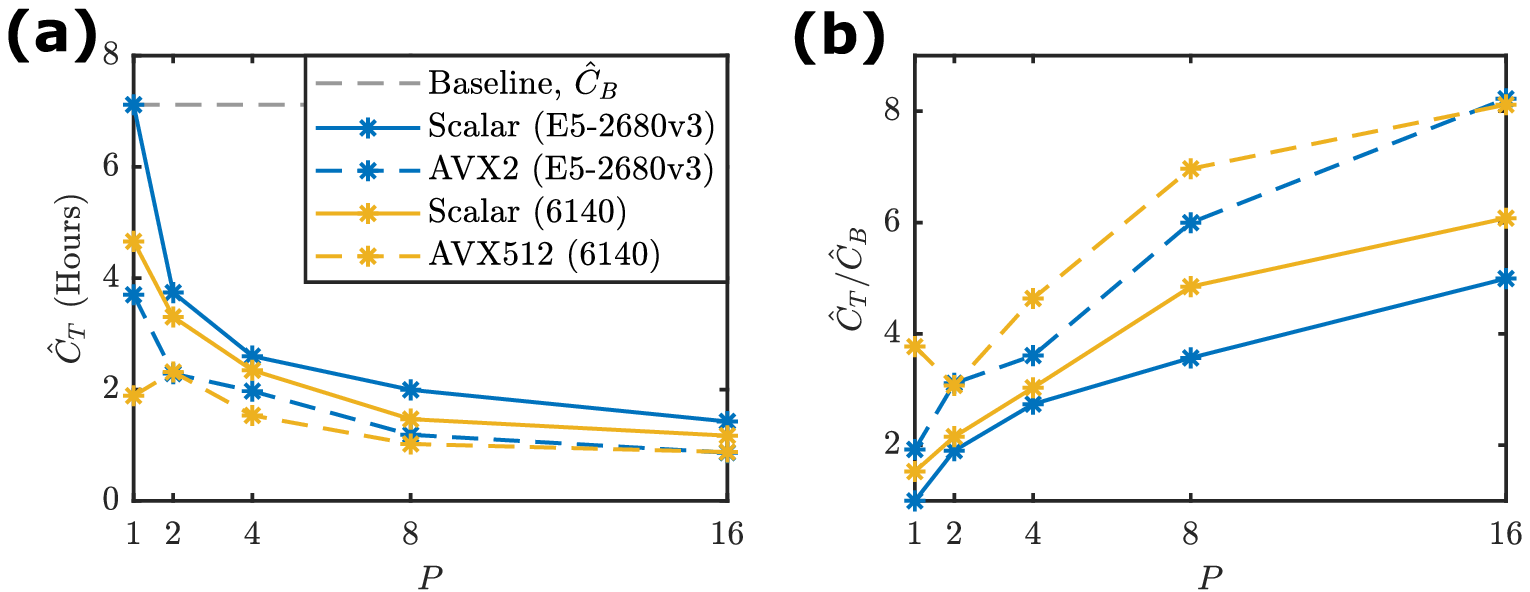}
		\caption{Benchmarking results for the SMC sampler for BEGE parameter inference. (a) Mean runtimes, $\hat{C}_T$, using $P$ cores for various implementations:  C with scalar operations only (Xeon E5-2680v3; blue solid), C with AVX2 vectors (Xeon E5-2680v3; blue dashed), C with scalar operations only (Xeon Gold 6140; yellow solid), and C with AVX512 vectors (Xeon Gold 6140; yellow dashed). (b) Computational improvements for each optimisation relative to baseline  $\hat{C}_B$ (C with scalar operations only with $P = 1$ on Xeon E5-2680v3).}
	\label{fig:bege}
\end{figure}

We implement the  BEGE inference problem using adaptive SMC with likelihood annealing with $N_p = 1024$ particles. We approximate $\CondCDFsub{u_t}{\sigma_p,\sigma_n,p_{t-1},n_{t-1}}{\text{BEGE}}$ using Equations~\eqref{eq:BEGE_LL_approx} and \eqref{eq:IGF_series_vec} with the  discretisation $\{\omega_{p,t}^j\}_{0 \leq j \leq N_\omega}$, $N_\omega = 100$, $\omega_{p,t}^0 = 10^{-4} -p_{t-1}\sigma_p$, and $\Delta \omega = (10\sigma_p\sqrt{p_{t-1}} -\omega_{p,t}^0)/(N_\omega-1)$, as is performed by~\cite{Bekaert2015} and \cite{South2019}. Results are provided in Figure~\ref{fig:bege} .

We observe an improvement of up to $2\times$ for AVX2 and almost $4\times$ for AVX512 regardless of the number of cores. This is improved VPU utilisation for AVX512 compared with the weak informativity test (Section~\ref{sec:wip}). That is, the BEGE distribution function approximation is a sufficiently large proportion of the total computation cost, that efficient calculation can exploit the longer 512 bit vectors.   
However, the overall speedup factor is lower from multithreading and increases slowly with $P$. The SMC sampler requires threads to synchronise at the end of each iteration. This synchronisation is in the resampling and annealing steps, and in estimating the optimal MCMC proposal scaling could be causing bottlenecks. This can be seen in the diminishing returns on the parallel speed-up as the number cores increases.     
\FloatBarrier
\subsubsection{Summary}
We have demonstrated how SIMD can be used to further accelerate a parallel SMC sampler for a challenging inference problem from econometrics. Note that the log-likelihood approximation we apply, based on the work of Bekaert et al. \cite{Bekaert2015} is biased. Recently, Li et al. \cite{Li2019} proposed an unbiased likelihood estimator, for which there are SIMD opportunities also. For the purposes of this manuscript, we find the biased approximation of~\cite{Bekaert2015} lends itself more direct discourse. 

\section{Discussion}
\label{sec:dis}

Across each of our example applications there have been some common features, which allow several conclusions to be drawn regarding task suitability for vectorisation. Firstly, it is clear that the form of stochastic model under study can have a dramatic effect on the potential performance boost due to vectorisation. Simulation schemes such as Euler-Maruyama are ideal candidates for vectorisation, however, exact stochastic simulation methods like the Gillespie direct method are more challenging. We show this effect in the supplementary material using the application of ABC to models of Tuberculosis transmission~\cite{Tanaka2006} and observe only moderate improvements from SIMD. This highlights a difference between continuous time and discrete time Markov processes, and may motivate the practitioner to consider alternative algorithms that are more suited to SIMD, for example, the $\tau$-leap approximations to the Gillespie algorithm. Similarily, the success of the SIMD version of the MCMC proposal kernel in Section \ref{sec:wip} relied on a SIMD implementation of the likelihood function. 

Secondly, each application involved nested parallelisation. The utility of SIMD here is that within each parallel thread, the computational tasks may be further sub-divided through use of VPUs. Efficiency gains due to SIMD are multiplicative to those arising from the use of thread parallelisation. Finally, each application involved a mixture of tasks that can be performed completely independently and tasks that require synchronisation or communication. This is important, since independent parallel computation is ideal for multithreading and fine grain parallel operations with frequent synchronisation are well suited for SIMD. Our demonstrations are widely applicable, since these three features are common to many Bayesian applications.  

Although we have focused on the acceleration of Monte Carlo methods for Bayesian applications, likelihood-based and frequentist applications can also benefit. This is especially true for complex optimisations for maximum likelihood parameter estimates~\cite{Hurn2016}. 
Furthermore, other Monte Carlo schemes can benefit from our approach. Any application that has been demonstrated using GPGPUs could exploit SIMD on the CPU~\cite{Lee2010,Holbrook2020}. This is rarely taken into consideration when comparison between GPGPUs and CPUs are made~\cite{Holbrook2020,Hurn2016,Lee2010b}.   

We have provided example C programs, using OpenMP and MKL libraries. We appreciate that, for very good reasons, higher level languages such as Matlab and R are often the preferred environments for many practitioners. The implementation techniques we present can be exploited through Matlab C-MEX or Rcpp interfaces. However, Matlab and R must be configured correctly to use the required compiler options. Of course, matrix operations performed within high level languages, such as Matlab or R, are likely already utilising SIMD via high-performance BLAS and LAPACK libraries.

Unfortunately, many of the SIMD and memory optimisations we presented cannot be directly exploited using a high level language alone. This is demonstrated practically in Section~\ref{sec:ABC}.
The only reliable way to access SIMD level parallelism from within a high level language is to use built-in matrix or vector functions that are already optimised (for example, by compiling R to use MKL). Memory access is still a significant challenge here. Between successive high level functions calls, it is unlikely that the caches are preserved, and as a result, our SIMD versions of the Euler-Maruyama scheme and MCMC proposal cannot be directly replicated in Matlab or R alone. A notable exception to this rule is the Julia language~\cite{Bezanson2017} using the \texttt{@simd} macro to achieve high performance. 

OpenMP is also not the only way to access SIMD within C/C++. For example, while OpenCL is primarily applied to GPCPUs and other accelerators, OpenCL kernels may be compiled for CPUs that support SIMD units~\cite{Hurn2016,Macintosh2019}. Instruction level intrinsic functions (also available in R via the RcppXsimd package) allow advanced features such as efficient random variates for small vectors, but this approach is very challenging and akin to machine code. 

%We have also presented realistic demonstrations in which Amdahl's law truly affects the maximum performance gains that are possible through vectorisation. This is not a limitation of SIMD, but is rather the nature of parallel algorithms. In fact, the algorithm types that suit vector processing cannot always be efficiently implemented with threads.

Many other Monte Carlo and Bayesian applications could benefit from these approaches. 
One possibility is large scale particle filters, perhaps operating within a pseudo-marginal scheme~\cite{Andrieu2010}, which could be implemented as a hybrid algorithm in which proposals and weight updates are performed in SIMD blocks that are distributed across multiple threads. Such a scheme would enable prefix summations, that is, parallel computation of cumulative sums, to be performed~\cite{Hillis:1986:DPA:7902.7903} in the resampling step, which may improve the performance.  
Another possibility is within the class of ABC validation or post-processing procedures. For example, in the recalibration post-processing technique of \cite{Rodrigues2018}, each of the $N$ samples $\paramvec^*$ from the approximate posterior distribution are individually recalibrated by the construction of a further approximate (ABC) posterior for each $\paramvec^*$ using all previously generated $(\paramvec^*,\mathcal{D}^*)$ pairs, but based on observing the associated (simulated) dataset ${\mathcal D}^*$, and constructing all univariate marginal posterior distribution functions. This re-use of ABC algorithm and previously generated parameter values and datasets is very common in ABC \cite[e.g.][]{Blum2013}, and makes them particularly suited for performance gains through parallelism and SIMD.

We have demonstrated that 
%In this work, we demonstrate, through practical and topical examples, techniques to maximise the utilisation of modern CPUs. In particular, we show that 
by following a few simple guidelines to maximise the utilisation of modern CPUs, advanced Monte Carlo methods may be relatively straightforwardly accelerated by a factor of up to $6 \times$ in addition to further speedups obtained through multithreading. 
These techniques will only become more relevant in the future as CPUs architectures are released with wider VPUs and statisticians develop more complex and sophisticated inferential algorithms.

\paragraph{Acknowledgements}
C.D. was supported by the Australian Research Council (ARC) under the Discovery Project scheme (DP200102101). S.A.S. was supported by the ARC under the Discovery Project scheme (DP160102544). C.D., S.A.S., and D.J.W. are supported by the Australian Centre of Excellence for Mathematical and Statistical Frontiers (ACEMS; CE140100049). C.D. and D.J.W. also acknowledge support frm the Centre for Data Science, Queensland University of Technology. Computational resources were provided by the eResearch Office, Queensland University of Technology.
%\end{acknowledgement}

%\begin{supplement}
%\sname{Supplement A}\label{suppA} 
%\stitle{Bayesian computations using SIMD operations}
\paragraph{Software availability}
Source code for the tutorial and case studies are freely available on GitHub 
\href{https://github.com/davidwarne/Bayesian_SIMD\_examples}{https://github.com/davidwarne/Bayesian\_SIMD\_examples}
%\sdescription{Contains: Tutorial code using R and C and example C code using OpenMP and  MKL. }
%\sname{Supplement B}\label{suppB} 
%\stitle{Runtime Tables}
%\slink[url]{https://github.com/davidwarne/Bayesian\_SIMD\_examples}
%\sdescription{Contains: Tutorial code using R and C and case study example C code using OpenMP and  MKL.}
%\end{supplement}

\newpage
\begin{appendices}
	\appendix
	
	\section{Benchmark Tables}
	\numberwithin{equation}{section}
	\numberwithin{figure}{section}
	\numberwithin{algorithm}{section}
	\numberwithin{table}{section}
	\setcounter{equation}{0}

Data tables of mean runtimes used for Figures 3--5 in the main text are provided in Tables~\ref{tab:pps},\ref{tab:wip} and \ref{tab:bege}

\begin{table}[h]	
	\caption{\small Computational performance R and C implementations of prior predictive sampling for the toggle switch model.}
	\begin{tabular}{c|rrrr|rr}
		\hline
		\multicolumn{1}{c}{Number of threads} & \multicolumn{4}{c}{E5-2680v3 Runtimes (Seconds)} & \multicolumn{2}{c}{6140 Runtimes (Seconds)}\\
		$P$ & R Na\"{i}ve  & R Optimise & C Scalar & C AVX2 & C Scalar & C AVX512 \\
		\hline
		1 & 298,041 & 10,502 & 2,632 & 577 & 2,395 & 380\\
		2 & 148,677 & 5,270 & 1,343 & 315 & 1,223 & 213\\
		4 & 76,520 & 2,663 & 760 & 179 & 631 & 128\\
		8 & 42,387 & 1,349 & 405 & 111 & 336 & 86\\
		16 & 16,961 & 681 & 223 & 78 & 190 & 69\\
		\hline
	\end{tabular}
	
	\label{tab:pps}
\end{table}

\begin{table}[h]	
	\caption{\small Computational performance for the weak informativity test. }
	\begin{tabular}{c|rr|rr}
		\hline
		\multicolumn{1}{c}{Number of threads} & \multicolumn{2}{c}{E5-2680v3 Runtimes (Seconds)} & \multicolumn{2}{c}{6140 Runtimes (Seconds)}\\
		$P$ & C Scalar & C AVX2 & C Scalar & C AVX512 \\
		\hline
		1 & 2,871 & 1,389 & 2,952 & 1,325\\
		2 & 1,429 & 701 & 1,463 & 665\\
		4 & 714 & 350 & 730 & 332\\
		8 & 359 & 178 & 367 & 167\\
		16 & 183 & 91 & 187 & 90\\
		\hline
	\end{tabular}
	
	\label{tab:wip}
\end{table}

\begin{table}[h]	
	\caption{\small Computational performance for BEGE parameter inference. }
	\begin{tabular}{c|rr|rr}
		\hline
		\multicolumn{1}{c}{Number of threads} & \multicolumn{2}{c}{E5-2680v3 Runtimes (Seconds)} & \multicolumn{2}{c}{6140 Runtimes (Seconds)}\\
		$P$ & C Scalar & C AVX2 & C Scalar & C AVX512 \\
		\hline
		1 & 25,618 & 13,327 & 16,780 & 6,793\\
		2 & 13,469 & 8,230 & 11,895 & 8,334\\
		4 & 9,358 & 7,097 & 8,446 & 5,526\\
		8 & 7,180 & 4,272 & 5,284 & 3,678\\
		16 & 5,131 & 3,115 & 4,215 & 3,156\\
		\hline
	\end{tabular}
	
	\label{tab:bege}
\end{table}
\newpage
\section{Tuberculosis transmission example}
The second model describes tuberculosis transmission dynamics~\cite{Tanaka2006,Sisson2007,Fearnhead2012}. The evolution of the number of tuberculosis cases over time, $N(t)$, is given by
\begin{equation*}
N(t) = \sum_{i=1}^{G(t)} X_i(t),
\end{equation*}
where $G(t)$ is the number of unique genotypes of the tuberculosis bacterium at time $t$ and $X_i(t)$ is the number of cases caused by the $i$th genotype. The time-varying evolution of the $X_i(t)$'s is driven by a linear birth-death process with birth rate $\alpha$ and death rate $\delta$. Mutation events occur at rate $\tau$ resulting in new genotypes and increases $G(t)$.

Given values for $\alpha$, $\delta$ and $\tau$, exact realisations can be generated using the \emph{Gillespie Direct Method}~\cite{Gillespie1977} with propensity functions,
\begin{equation*}
a_b(X_i(t)) = \alpha X_i(t), \quad
a_d(X_i(t)) = \delta X_i(t), \quad
a_m(X_i(t)) = \tau X_i(t),
\end{equation*}
where $a_b$, $a_d$ and $a_m$ are the birth, death and mutation propensities respectively.  

\cite{Tanaka2006} demonstrate the application of ABC methods to infer the parameters $\theta=(\alpha, \delta,\tau)'$ for a real dataset of tuberculosis cases in San Francisco during the early 1990s that consists of $326$ distinct genotypes. The number of distinct genotypes and the genetic diversity are used as summary statistics (see \cite{Tanaka2006}).

\subsection{Parallelisation and vectorisation opportunities}

Given $P$ cores, the process of i.i.d.~sampling $(\theta^*,{\mathcal D}^*)$ from the joint prior predictive can be trivially divided between cores. For the toggle switch model (Equation~\eqref{eq:gene_uv_SDE}) it is reasonable to divide the workload evenly since the number of floating point operations per realisation is constant. This is called a \texttt{static} schedule for \texttt{OpenMP}. Each core will generate $N_P = N/P$ prior predictive samples. We assume $N_P$ is an integer.

For the tuberculosis model, generation of the prior predictive samples involves a discrete-state, continuous-time Markov process that is implemented with the Gillespie algorithm. The simulation times can vary wildly and a \texttt{static} schedule can result in a load imbalances. Therefore, \texttt{guided} or \texttt{dynamic} scheduling could perform better. In practice, we find the overhead of load imbalance is relatively low for large $N$. Furthermore, an early termination rule, as used in the Lazy ABC approach of~\cite{Prangle2016} would further reduce imbalances. We conclude that \texttt{static} distribution of prior predictive samples is appropriate as a general strategy for ABC applications. Reproducibility of RNG sequences is also possible with \texttt{static} scheduling. 

No such general strategy exists for vectorisation. The simulation algorithms for the two models are very different. The Euler-Maruyama scheme is applied to toggle switch model and the Gillespie Direct method is applied to the tuberculosis model. The performance improvement opportunity naturally differs. 

Since each Euler-Maruyama simulation performs identical operations for each cell $i =~1,2\ldots,C$, this is ideal for vectorisation. We evolve cells in blocks of length $V$ where $64V$ is the bit width of the VPU inputs. Aside from the generation of Gaussian random variates, every operation is replaced with SIMD operation. We generate $2n_tV$ Gaussian variates using the Intel MKL RNG ahead of processing each block. This is efficient when $n_t V$ is within L1 cache capacity, but large enough to take advantage of the Intel routines. Typically, the Intel routines require at least $1,000$ variates to be generated for  peak performance.

Fewer opportunities exist for vectorisation of the Gillespie method used for the tuberculosis model. The only suitable vectorisable step for updating genotype weights for the lookup method to select the genotype to apply the next event. The discrepancy measure components, $g$ and $H$, can also be vectorised. The length of the state vector in the tuberculosis model is large enough for some performance boost from the vectorisation to be noticeable. The genetic diversity $H$ calculation is particularly efficient as the sum of squares operation can exploit fused-multiply-add (FMA) SIMD operations that performs $A \leftarrow A + (B\circ C)$.

\subsection{Performance}
We report performance improvements obtained through vectorisation and multithreading for two CPU architectures, the Xeon E5-2680v3 (Haswell)\footnote{\href{https://ark.intel.com/products/81908/Intel-Xeon-Processor-E5-2680-v3-30M-Cache-2-50-GHz-}{https://ark.intel.com/products/81908/Intel-Xeon-Processor-E5-2680-v3-30M-Cache-2-50-GHz-}} and Xeon Gold 6140 (Skylake)\footnote{\href{https://ark.intel.com/products/120485/Intel-Xeon-Gold-6140-Processor-24-75M-Cache-2-30-GHz-}{https://ark.intel.com/products/120485/Intel-Xeon-Gold-6140-Processor-24-75M-Cache-2-30-GHz-}}. The Xeon E5-2680v3 supports Intel's AVX2 instruction set ($256$ bit vectors), and the Xeon Gold 6140 supports Intel's AVX512 instruction set ($512$ bit vectors). %The Xeon Gold series is a part of Intel's latest Scalable processor family that inherits much of the VPU technology from the Intel Xeon Phi family.

Tables~\ref{tab:tb_avx256} and \ref{tab:tb_avx512} show the results for the tuberculosis model for a range of prior predictive sample counts, $N$. The speedup factor from vectorisation is unsurprisingly not as significant as for the toggle switch model. This demonstrates the limitations of SIMD operations (including GPGPU-based accelerators). There is still an improvement from vectorisation of between $1.4 \times$ to $1.8 \times$. A good result in the context of the HPC literature~\cite{Dongarra2014,Kozlov2014,Mudalige2016}. 
\begin{table}	
	
	\begin{tabular}{crrrr}
		\hline
		\multicolumn{1}{c}{Number of samples} & \multicolumn{4}{c}{Runtime (Seconds) [Speedup factor]} \\ %\multicolumn{2}{c}{MCMC-ABC}
		$N$ & sequential & sequential+SIMD & parallel & parallel+SIMD \\
		\hline
		$500$ & $46.7$ [$1.0 \times$] & $25.5$ [$1.8 \times$] & $64.5$ [$0.7 \times$] & $36.0$ [$1.3 \times$] \\
		$1,000$ & $230.2$ [$1.0 \times$] & $127.7$ [$1.8 \times$] & $72.1$ [$3.2 \times$] & $40.1$ [$5.8 \times$]\\ 
		$2,000$ & $385.0$ [$1.0 \times$] & $212.7$ [$1.8 \times$] & $91.8$ [$4.2 \times$] & $50.8$ [$7.6 \times$] \\
		$4,000$ & $657.4$ [$1.0 \times$] & $363.1$ [$1.8 \times$] & $229.5$ [$2.9 \times$] & $127.7$ [$5.2 \times$] \\
		$8,000$ & $1,175.6$ [$1.0 \times$] & $649.2$ [$1.8 \times$] & $386.7$ [$3.0 \times$] & $212.5$ [$5.5 \times$] \\
		$16,000$ & $2,601.7$ [$1.0 \times$] & $1,449.4$ [$1.8 \times$] & $697.3$ [$3.7 \times$] & $389.6$ [$6.7 \times$] \\
		\hline
	\end{tabular}
	\caption{\small Computational performance of computing prior predictive samples under the tuberculosis model using an Intel Xeon E5-2680v3 (Haswell) processor. Parallel times are based on $4$-way multithreading and SIMD times using $256$ bit vector operations.}	
	\label{tab:tb_avx256}
\end{table}

\begin{table}	
	\begin{tabular}{crrrr}
		\hline
		\multicolumn{1}{c}{Number of samples} & \multicolumn{4}{c}{Runtime (Seconds) [Speedup factor]} \\ %\multicolumn{2}{c}{MCMC-ABC}
		$N$ & sequential & sequential+SIMD & parallel & parallel+SIMD \\
		\hline
		$500$ & $15.9$ [$1.0 \times$] & $11.5$ [$1.4 \times$] & $23.7$ [$0.7 \times$] & $17.9$ [$0.9 \times$] \\
		$1,000$ & $83.9$ [$1.0 \times$] & $61.5$ [$1.4 \times$] & $26.3$ [$3.2 \times$] & $19.3$ [$4.4 \times$] \\
		$2,000$ & $138.2$ [$1.0 \times$] & $101.2$ [$1.4 \times$] & $32.8$ [$4.2 \times$] & $24.0$ [$5.8 \times$] \\
		$4,000$ & $233.5$ [$1.0 \times$] & $170.7$ [$1.4 \times$] & $83.8$ [$2.8 \times$] & $61.5$ [$3.8 \times$] \\
		$8,000$ & $412.9$ [$1.0 \times$] & $301.9$ [$1.4 \times$] & $138.1$ [$3.0 \times$] & $101.1$ [$4.1 \times$] \\
		$16,000$ & $922.5$ [$1.0 \times$] & $675.2$ [$1.4 \times$] & $249.7$ [$3.7 \times$] & $183.0$ [$5.0 \times$] \\
		\hline
	\end{tabular}
	\caption{\small Computational performance of computing prior predictive samples under the tuberculosis model using an Intel Xeon Gold 6140 (Skylake) processor. Parallel times are based on $4$-way multithreading and SIMD times using $512$ bit vector operations.}	
	\label{tab:tb_avx512}
\end{table}

The speedup factor due to vectorisation is larger, at $1.8 \times$, for the Xeon E5-2680v3 compared with $1.4\times$ for the Xeon Gold 6140. However, the Xeon Gold 6140 results demonstrate almost $3 \times$ improvement in the scalar performance over the Xeon E5-2680v3. Therefore, we conclude that the tuberculosis model simulation is accessing L3 cache and main memory more often, and thus taking advant memory bandwidth and extra memory channels on the Xeon Gold 6140. If the tuberculosis model is not utilising L1 and L2 cache as intensively, then this explains the reduced speedup for wider vectors as the VPUs is not efficiently utilised.

Differences in performance of the toggle switch model and the tuberculosis model suggest the choice of stochastic simulation algorithm is crucial. Introducing approximations \cite{Gillespie2000,Gillespie2001} for the tuberculosis model could significantly improve performance. 

\end{appendices}

\end{document}